\documentclass[aps,prl]{revtex4}
\usepackage{amsfonts}          
\usepackage{amssymb}           
\usepackage{amsmath}   

\usepackage{graphics}
\usepackage{epstopdf, epsfig}

\usepackage{amsmath}
\DeclareMathAlphabet{\pazocal}{OMS}{zplm}{m}{n}

\newcommand{\calQ}{\pazocal{Q}}
\newcommand{\calW}{\pazocal{W}}
\newcommand*{\rmd}{{\mathrm d}}

\newcommand*{\bfe}{{\bf e}}
\newcommand*{\bfn}{{\bf n}}
\newcommand*{\bfu}{{\bf u}}

\newcommand*{\bfw}{{\bf w}}
\newcommand*{\bfx}{{\bf x}}
\newcommand{\be}{\begin{equation}}
\newcommand{\ee}{\end{equation}}
\newcommand{\bea}{\begin{eqnarray}}
\newcommand{\eea}{\end{eqnarray}}
\newcommand{\tn}{\textnormal}
\newcommand{\tthird}{\textstyle{\frac{1}{3}} \displaystyle }

\newcommand{\tfourth}{\textstyle{\frac{1}{4}} \displaystyle }

\newcommand{\twentysixthirds}{\textstyle{\frac{26}{3}} \displaystyle }
               
\newcommand{\bom}{\mbox{\boldmath $\omega$}}
\newcommand\thalf{\ensuremath{{\textstyle\frac{1}{2}}}}

\usepackage{xcolor}

\def\today{
\number\day\space
\ifcase\month\or
January\or February\or March\or April\or May\or June\or
July\or August\or September\or October\or November\or December\fi
\space\number\year}

\begin{document}

\title{Flow induced by the rotation of two circular cylinders in a viscous fluid.}

\author{E. Dormy$^{1}$ and H.K. Moffatt$^{2}$}
\affiliation{$^1$D\'epartement de Math\'ematiques et Applications,
  UMR-8553, \'Ecole Normale Sup\'erieure, CNRS, PSL University, 75005
  Paris, France\\
$^2$Department of Applied Mathematics and Theoretical
  Physics, Wilberforce Road,\\ Cambridge CB3 0WA, UK
}

\date{\today}


\begin{abstract}
The low-Reynolds-number Stokes flow driven by rotation of two parallel
cylinders of equal unit radius is investigated by both  analytical and
numerical techniques. In Part I, the case of counter-rotating cylinders is
considered.  A numerical (finite-element) solution is obtained by enclosing
the system in an outer cylinder of radius $R_{0}\!\gg\!1$, on which the
no-slip condition is imposed.  A model problem  with the same symmetries is
first solved exactly, and the limit of validity of the Stokes approximation
is determined; this model has some relevance for ciliary propulsion.  For
the two-cylinder problem, attention is focused on the small-gap situation
$\varepsilon \ll 1$. An exact analytic solution is obtained in the contact
limit $\varepsilon=0$, and  a net force  $F_{c}$ acting on the
pair of cylinders in this contact limit
is identified; this contributes to the torque that each cylinder experiences about its axis. The far-field torque doublet (`torquelet') is also identified.

Part II treats the case of co-rotating cylinders, for which again a finite-element numerical solution is obtained for $R_{0}\!\gg \!1$.   The theory of Watson (1995, {\em Mathematika}, {\bf 42}, 105--126)  is elucidated and shown to agree well with the numerical solution. In contrast to the counter-rotating case, inertia effects are negligible throughout the fluid domain, however large, provided Re $\ll 1$.

In the concluding section,  the main results for both cases are summarised, and the situation when the fluid is unbounded ($R_{0}=\infty$) is discussed. If the cylinders are free to move (while rotating about their axes), in the counter-rotating case they will then translate relative to the fluid at infinity with constant velocity, the drag force exactly compensating the self-induced force due to the counter-rotation.  In the co-rotating case, if the cylinders are free to move, they will rotate as a pair  relative to the fluid at infinity and the net torque on the cylinder pair is zero;  the flow relative to the fluid at infinity is identified as a `radial quadrupole'.  If, on the other hand,  the cylinder axes are held fixed, the Stokes flow in the counter-rotating case extends only for a distance $r\sim \tn{Re}^{-1}\log{[\tn{Re}^{-1}]}$ from the cylinders, and it is argued that the cylinders then experience a (dimensionless) force $\hat{F}_{y}\sim 1/\log{[\tn{Re}^{-1}\log{[\tn{Re}^{-1}]}]}$; in the co-rotating case, the cylinder pair experiences a (dimensionless) torque $\hat{\mathcal{T}}$, which tends to 17.2587 as $\varepsilon \downarrow 0$; this torque is associated with a vortex-type flow $\sim r^{-1}$ that is established in the far field. 

 Situations that can be described by the condition $\varepsilon < 0 $ are treated for both counter- and co-rotating cases in the supplementary material. 
\end{abstract}

\maketitle

\vskip 2mm
\noindent {\bf Introduction}
\vskip 2mm
\noindent
The very classical problem of determining the two-dimensional flow induced by the rotation of two adjacent cylinders of equal unit radii originated with the pioneering work of Jeffery (1915, 1922) \cite{J15} \cite{J22}.  Jeffery's solution involved a velocity field that failed to tend to zero at infinity (`Jeffery's paradox'). Our motivation in the present study is to understand how this paradox may be resolved.  The cases of counter-rotating or co-rotating cylinders are very different and are treated separately in Parts I and II below.  In both cases, we circumvent the condition of `zero velocity at infinity' by enclosing the fluid in a cylinder  of large radius $R_{0}\gg 1$, on which a no-slip condition may be imposed.  We then seek to infer the behaviour in the limiting situation when the fluid is unbounded ($R_{0}=\infty$).  Important applications are to be found in the theory of swimming microorganisms as treated in the book of Lauga (2020) \cite{L20}, or in the behaviour and interaction of two-dimensional microswimmers, as studied recently by Nie et al (2023) \cite {N2023}. 

We pay particular attention to the situation when the gap $2\varepsilon$ between the cylinders is small.  The combination of conditions $R_{0}\gg 1$ and $\varepsilon\ll 1$ is numerically challenging, but we can exploit lubrication theory in the small gap and check the accuracy and range of applicability of the finite-element numerical scheme that is adopted.  The problem studied here is prototypical, in that our approach may be extended to a range of problems in either two or three dimensions where the topology is non-trivial and small gaps (or contact) between moving solid bodies may occur.

\section{Part I: counter-rotating cylinders}
\begin{figure}
\begin{center}
\includegraphics*[width=0.7\textwidth,  trim=0mm 0mm 0mm 0mm]{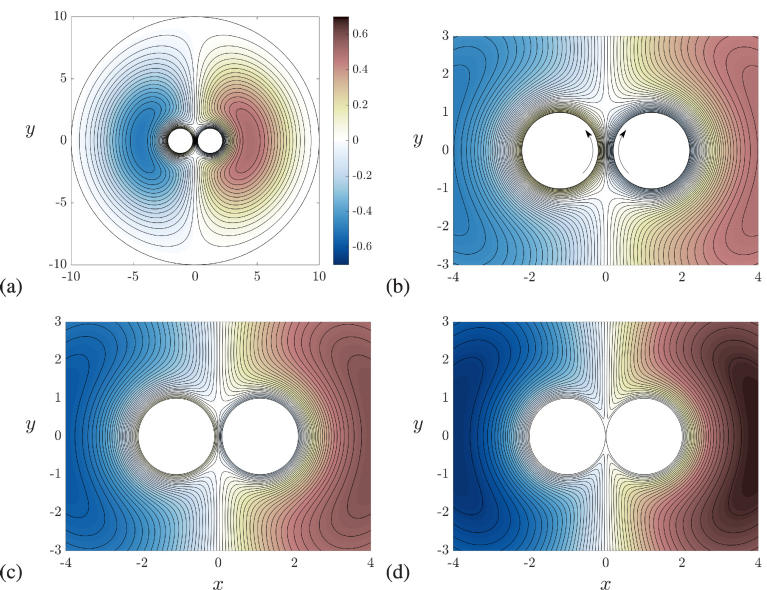} 
\end{center}
\caption{Streamlines $\psi=$ const.~(symmetric with respect to both $x$-axis and $y$-axis) from the numerical solution of the Stokes problem for counter-rotating cylinders; the outer boundary condition is no-slip on $r=R_{0}$ (here $R_{0}=10$ and $\psi=0$ on $r=R_{0}$);  the colour code is the same for all four panels. (a) $\varepsilon=0.2$, full flow domain;  (b) $\varepsilon=0.2$, zoom to neighbourhood of  inner cylinders; the sense of rotation is indicated by the arrows; (c) same zoom for $\varepsilon=0.1$; (d) same zoom for $\varepsilon=0.01$. Note that the two saddle points on the 
$y$-axis approach the origin as $\varepsilon$  decreases.}
 \label{4pannels_num}
\end{figure}

\vskip 2mm
\noindent {\bf 1. Background}
\vskip 2mm
\noindent  The problem of determining the two-dimensional steady Stokes flow induced by the counter-rotation of two parallel cylinders of infinite length in an unbounded viscous fluid was initiated in the seminal work of Jeffery (1922) \cite{J22}.The symmetries of the streamline patterns (the angular velocities being equal and opposite) are as indicated in figure \ref{4pannels_num}.  Using bipolar coordinates and conformal transformation, Jeffery showed that it was not possible to satisfy the condition that the fluid velocity should vanish at infinity (`Jeffery's paradox'). He did however find a solution that allowed for a uniform streaming flow at infinity. [The situation is in some respects like the classical problem of viscous flow past a circular cylinder, for which Stokes had himself established that there is no solution of the linearised equations satisfying the required condition at infinity  \cite{L32}, \cite{B67};  this difficulty was ultimately resolved by Proudman \& Pearson \cite{PP57} through the (then novel) technique of matched asymptotic expansions]. For ease of reference, some aspects of Jeffery's approach are summarised in Appendix A, together with  a new asymptotic evaluation of his results.

We use dimensionless variables such that the two cylinders $\mathcal{C}_{1,2}$ have equal unit radius $a=1$,  are centred at $(\mp(1+\varepsilon),0)$, and rotate with angular velocities  $\Omega_{1}=1$, $\Omega_{2}=-1$, respectively; the no-slip condition is to be satisfied on both cylinders.  The natural Reynolds number for the problem is then 
\be
\tn{Re} =\Omega_{1} a^{2}/\nu=\nu^{-1}\,,
\ee
 where $\nu$ is the kinematic viscosity in dimensionless units.  The gap between the cylinders is $2\varepsilon$, and their boundaries are 
\be\label{cyl_bdy}
\mathcal{C}_{1}: (x+1+\varepsilon)^2 +y^2=1 \quad \tn{and}\quad \mathcal{C}_{2}: (x-1-\varepsilon)^2 +y^2 =1\,.
\ee
We shall be particularly concerned with the `small-gap situation', $0<\varepsilon\ll 1$. In the  limiting situation $\varepsilon = 0$,  the cylinders make contact and a singularity is to be expected.  [The seemingly artificial situation of `overlapping cylinders' ($\varepsilon < 0$), which could in principle be realised by ciliary action of micro-organisms, is considered in the supplementary material.]

In the Stokes approximation, valid at small Reynolds number, the  pressure field $p$ and velocity field $\bfu =(u,v)= (\partial \psi/\partial y, -\partial\psi/\partial x)$  satisfy the equations
\begin{equation} 
 \mu \nabla^2 \bfu = \nabla p \, ,\quad \nabla \cdot \bfu = 0 \, ,
\label{Stokes2}
\end{equation}
where $\mu$ is the dynamic viscosity.  The vorticity $\bom=\nabla\times\bfu=-\nabla^{2}\psi \,\,{\bf e}_{z}$ satisfies $\nabla^{2}\bom =0$, so that $\psi$  satisfies the biharmonic equation $ \nabla ^4 \psi = 0$ .

We shall  suppose that the fluid is bounded externally by a cylinder $\mathcal{C}_{0}$ of radius $r=R_{0}\gg 1$, on which an appropriate boundary condition can be imposed; in this way, we circumvent the problem of imposing conditions `at infinity'. This  is moreover needed for the numerical treatment  presented in this paper, and is in any case more realistic in experimental contexts. We note the comprehensive numerical and experimental investigations by Hills(2002) \cite{H02}  of the `two-roll mill' flow in a finite domain with a rectangular outer boundary. Hills used finite-difference techniques with focus on the effects of increasing Reynolds number and of change in the angular velocities of the cylinders.  
 
The two-cylinder problem was revisited by Watson(1995) \cite{W95} who addressed Jeffery's paradox, recognising that there may be a self-induced  force  per unit axial length on the pair of cylinders resulting from their counter-rotation. Such a force must generate  a two-dimensional stokeslet in the far flow field.  Watson argued that the resulting flow should be matched asymptotically to an outer solution of the full Navier-Stokes equations, presumably a jet of the type first analysed by Bickley (1937) \cite{B37}, but  this difficult matching problem remains unsolved to the present day. 

The  problem had been discussed earlier by Dorrepaal et al.(1984) \cite{D84}, who  showed on the basis of Jeffery's (1922) solution that the force on each cylinder is zero provided Jeffery's uniform streaming at infinity is  allowed for.  The problem was further investigated by Elliott et al.(1995a,b)  \cite{E95a,E95b} who used the boundary-element method to determine the flow for several choices of cylinder radii and angular velocities, with the conclusion that in general ``the total force and the total torque [on the pair of cylinders] are both zero".  This conclusion was at some variance with Watson's conclusions, published almost simultaneously. 

A further approach to the two-cylinder problem was adopted by Ueda et al.(2003) \cite{U2003}, who supposed that the two cylinders are abruptly set in their rotary motion at time $t=0$ in an initially quiescent fluid.  Neglecting non-linear inertia effects 
(i.e.~for vanishingly small Reynolds number) and  again using the boundary-element method, they studied the approach to a steady state as $t\rightarrow \infty$, and found expressions for the asymptotic force acting on each cylinder.  They found a non-zero force (figure 6b of their paper) in apparent conflict with the assertion of Elliott et al.~cited above. We note that a non-zero force implies a far-field stokeslet velocity $\sim \log{r}$,  incompatible with the outer boundary condition imposed for all $t$ by Ueda et al., suggesting that a steady state is not attained throughout the infinite domain.  The situation is indeed confusing, and calls for clarification, which we attempt to provide in the present paper.

\vskip 5mm
\noindent {\bf 2. A simplified model}
\vskip 2mm
\begin{figure}
\begin{center}
(a)\includegraphics*[width=.4\textwidth, trim=10mm 0mm 0mm 0mm, clip]{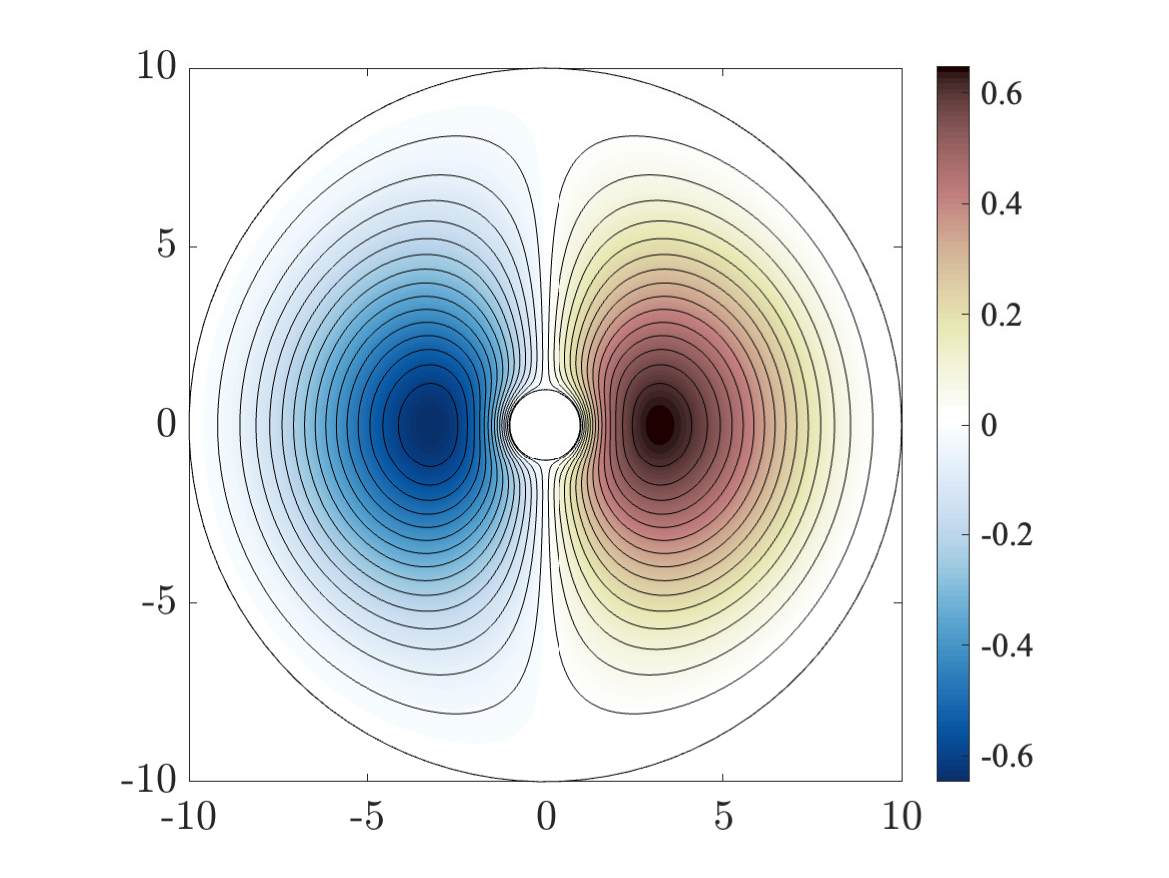} \quad
(b)\includegraphics*[width=.4\textwidth, trim=10mm 0mm 0mm 0mm, clip]{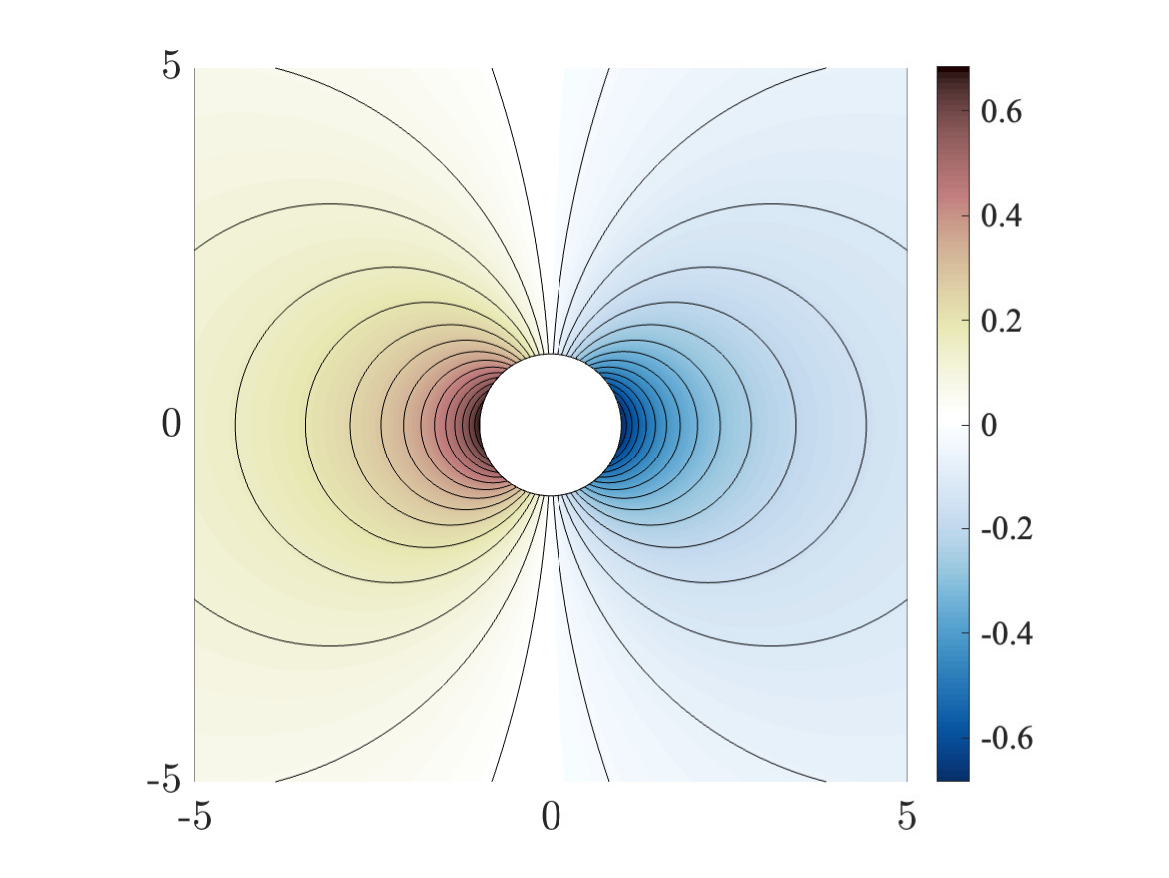} 
\end{center}
\caption{Streamlines $\psi(r,\theta)=$ const. for the flow (\ref{gen_soln}); (a) with no slip on  $r=R_{0}\, (\tn{here} \, R_{0} =10)$;  (b) with no outer boundary condition ($C=D=0$), and with $A=0$ thus showing the instantaneous (dipole) flow in a frame fixed to the fluid at infinity.}
\label{C_contours}
\end{figure}
\noindent It is helpful first to consider the following idealised problem [see figure \ref{C_contours}(a)]. Suppose that  fluid is contained in the annulus between two cylinders $r=1$ and $r=R_{0}$, and that the boundary conditions on the inner cylinder are
\be\label{bcs_on_1}
u_{r}(r,\theta)\equiv r^{-1}\partial\psi/\partial\theta=0,\quad u_{\theta}(r,\theta)\equiv -\partial\psi/\partial r=-\cos\theta\quad \tn{on}\,\,r=1\,.
\ee
Thus we replace the two counter-rotating cylinders by a single cylinder with prescribed tangential velocity on its surface --- a `squirming cylinder' in the language of bio-fluid mechanics \cite{L52,LP09}.  This problem has the same symmetries as the two-cylinder situation, and may be expected to give similar qualitative behaviour for large $R_{0}$. Again the Reynolds number based on the boundary conditions (\ref{bcs_on_1}) is simply Re $=\nu^{-1}$.

The general solution of $\nabla^{4}\psi =0$ proportional to  $\cos\theta$ is 
\be\label{gen_soln} 
\psi (r,\theta)= (A \, r + B \,r^{-1} + C\,r \log r +D\, r^{3})\cos\theta\,.
\ee
Here, the stokeslet term $ C\,r \log {r} \cos\theta$ is associated with the force 
\be\label{force}
F= -4\pi \mu C, 
\ee
 that the cylinder exerts on the fluid (of dynamic viscosity $\mu$).
The velocity components are
\be
u_{r}(r,\theta)=- (A + B\,r^{-2} + C\, \log r +D\, r^{2})\sin\theta\,,\quad
u_{\theta}(r,\theta)= -(A-B\, r^{-2}+C (1+\log r)+3D\,r^{2})\cos\theta\,,
\ee
and the boundary conditions (\ref{bcs_on_1}) give
\be \label{bcsABCD}
A+B+D =0,\quad A-B +C +3D=1\,.
\ee
The no-slip conditions $u_{r}=u_{\theta}=0$ on $r=R_{0}$  give the additional  equations 
\be\label{no_slip_R}
A + B\,R_{0}^{-2} + C\, \log R_{0} +D\, R_{0}^{2}=0,\quad A-B\,R_{0}^{-2}+C (1+\log R_{0})+3D\,R_{0}^{2}=0\,.
\ee
Equations (\ref{bcsABCD}) and (\ref{no_slip_R}) may be solved for the constants $A,B,C,D$, giving 
\be
A=\frac{ R_{0}^4(2 \log{R_{0}}\!- \!1)\! +\! 2  \log{R_{0}}\!+\!1}{4(R_{0}^2\!-\!1) H(R_{0})}\,,\,\,
B=\frac{-[R_{0}^4(2\log{R_{0}}\!-\!1)\!+\!R_{0}^2]}{4(R_{0}^2\!-\!1)H(R_{0})}\,,\,\,
C=\frac{1\!-\!R_{0}^2}{2H(R_{0})}\,,\,\,D=-(A+B)\,,
\ee
where $H(R_{0})= R_{0}^2 ( \log{R_{0}}\!-\!1)\!+\!  \log{R_{0}}\!+\!1$.
\vskip 2mm
For $\log{R_{0}}\gg 1$,  $C\sim -(2\log{R_{0}})^{-1}$, so from (\ref{force}) the force $F$ exerted by the cylinder on the fluid behaves like
\be 
F\sim 2\pi\mu/\log{R_{0}}\quad\tn{as}\,\,\log{R_{0}}\rightarrow\infty\,,
\ee
with very slow  convergence to zero.  The streamlines for this flow are shown in figure \ref{C_contours}(a) for the choice  $R_{0}=10$.

On $\theta=\pi/2$ and in the `inner regime'  $1\ll r\ll R_{0}$, the radial velocity is given asymptotically  by
\be\label{radial_asymptotic}
u_{r}(r,\pi/2)\sim -\frac{1}{2}+\frac{2\log{r}-1}{4\log{R_{0}}}+\tn{O}\left(\frac{1}{r^{2} \log{R_{0}}},\frac{r^2}{R_{0}^2 \log{R_{0}}}\right)\,,
\ee
the three terms representing (in order) contributions from a uniform stream $-1/2$, the stokeslet, and a (negligible) dipole. For $1\ll r\ll \log{R_{0}}$, the uniform stream is dominant; thus, for any fixed $r\gg 1$, the velocity tends to the  uniform-stream value $-1/2$ as $\log{R_{0}}\rightarrow\infty$. 

If instead we focus on the `outer regime' where  $r=\tn{O}(R_{0})$ as $R_{0}\rightarrow\infty$, we may set $k=r/R_{0}$ with $\{k<1, k=\tn{O}(1)$\}; then 
\be
u_{r}(r,\pi/2)\sim \frac{k^{2}-2\log{k}-1}{4(\log{R_{0}}-1)} \quad \tn{for}\,\, R_{0}\gg 1,\,\, r=\tn{O}(R_{0})\,.
\ee
The ratio of inertia forces to viscous forces in this regime is given by
\be\label{maxratio}
\frac{r \,u_{r}(r,\pi/2)}{\nu} =\frac{f(k)R_{0}}{\nu(\log{R_{0}}-1)}\,,\quad\tn{where}\,\,f(k)= \tfourth k(k^{2}\!-\!2\log{k}\!-\!1)\,.
\ee
The function $f(k)$ is maximal at $k\approx 0.244$, where it takes the approximate value 0.115.  Hence, neglecting constants of order unity, the maximum of the expression (\ref{maxratio}) is of order $\nu^{-1} R_{0}/ \log{R_{0}}$ for $\log{R_{0}}\gg 1$.  This means that if 
\be\label{R0_estimate}
\frac{R_{0}}{\log{R_{0}}}\gtrsim  \, \nu = \, \tn{Re}^{-1},\,\,\tn{or equivalently if}\,\,\,\, R_{0}\gtrsim \, \tn{Re}^{-1}\log{[ \tn{Re}^{-1}]}\,,
\ee
 then inertia cannot be neglected in this outer regime.   Thus, no matter how small Re may be, there is always an upper limit for $R_{0}$ for applicability of the Stokes approximation throughout the whole fluid domain $1<r<R_{0}$.  
 
For the same reason, in the case of an unbounded fluid domain ($R_{0}=\infty$), inertia effects must intervene when 
\be\label{inertia_limit}
r\gtrsim  \tn{Re}^{-1}\log{[ \tn{Re}^{-1}]}.
\ee 
The same limitation must surely apply to the two-cylinder problem, supporting the suggestion of Watson (1995) \cite{W95} that matching to a solution of the full Navier-Stokes equation is needed in this outer regime.

\vskip 2mm
\noindent {\em No outer boundary}
\vskip 1mm
\noindent We may note nevertheless that, if we take $C=D=0$ in
(\ref{gen_soln}), thus eliminating the  terms that are most divergent for
large $r$, then we have simply $A=-B=\thalf$, and the flow is a simple
potential flow (and so an exact solution of the Navier-Stokes equation)
with the uniform stream $[0,\,-1/2]$ at infinity.  Figure
\ref{C_contours}(b) shows the instantaneous streamlines in a frame of 
reference fixed to the fluid at infinity; this is a dipole field.  It is
interesting that this `swimming flow' could be generated by ciliary action
on the surface of a microscopic organism;  in contrast to the conventional
Stokes flow past a cylinder,  the force here evaluates to zero, because the
force generated by the ciliary action (\ref{bcs_on_1}) is exactly balanced
by the resulting drag force on the organism.  [A similar mechanism for a
  spherical microscopic swimmer has been described by Lauga \& Powers (2009) \cite{LP09}.]
\vskip 5mm \noindent
{\bf  3. The two-cylinder problem}
\subsubsection{3.1 Numerical solution}\label{Num_soln}
\noindent Consider now the two-cylinder counter-rotating situation with non-dimensionalised boundary conditions
\be
\Omega_{1}=1 \,\, \tn{on}\,\, \mathcal{C}_{1},\quad\Omega_{2}=-1 \,\, \tn{on}\,\, \mathcal{C}_{2},\quad\bfu=0\,\,\tn{on} \,\,\mathcal{C}_{0}\,(r=R_{0})\,.
\ee
We used a finite-element procedure (details in Appendix C)  to provide an accurate numerical solution.
\begin{figure}
\begin{center}
\includegraphics*[width=0.8\textwidth,  trim=0mm 0mm 0mm 0mm]{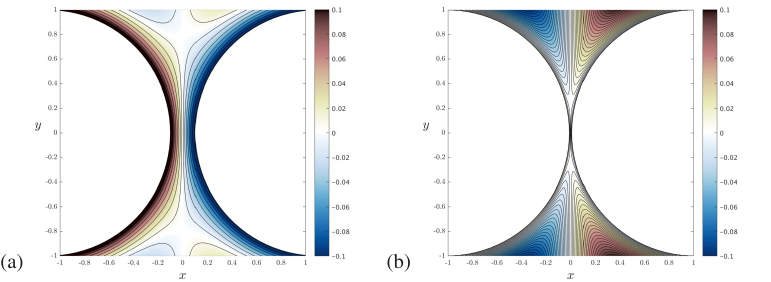} 
\end{center}
\caption{As for figure \ref{4pannels_num}(c,d),  further zoomed to near the origin; (a)  $\varepsilon = 0.1$; (b)
  $\varepsilon = 0.01 .$} 
  \label{2pannels_num}
\end{figure}
Figure \ref{4pannels_num}(a) shows the streamlines $\psi=$ const.~for the flow when  $\varepsilon=0.2$ and $R_{0} =10$; the arbitrary additive constant in $\psi$ was chosen so that $\psi=0$ on $r=R_{0}$.   Note the presence of two saddle points near the inner cylinders on the $y$-axis; these  are more evident in the zoom of figure \ref{4pannels_num}(b). There is an upward flux in the gap between the cylinders, and  the flow due to the rotation of each cylinder  exerts an upward force on the other.  This joint force is transmitted to the fluid, creating a stokeslet contribution to the flow beyond the cylinders, the possibility recognised by Watson (1995);  this is coupled with the downward streaming previously found by Jeffery (1922)  [cf eqn.~(\ref{radial_asymptotic})]. 

Figure \ref{4pannels_num}(c) shows the same zoom when $\varepsilon$ is reduced to $0.1$.   Here  the saddle points have moved towards the origin, a process that is further marked in  figure \ref{4pannels_num}(d) for which $\varepsilon=0.01$. The  zoom of figure \ref{2pannels_num}(a) to the neighbourhood of the origin for the case  $\varepsilon=0.1$ makes this even more evident;  figure  \ref{2pannels_num}(b) for $\varepsilon=0.01$ shows that the saddle points are now well within the narrow gap where lubrication theory should be relevant.  We shall find [see comment below equation (\ref{vax})] that the saddle points are located at $y=\pm (6\varepsilon)^{1/2}$ as $\varepsilon \rightarrow 0$.

\subsubsection{3.2 Lubrication theory}\label{lubrication_theory}
\noindent If we assume that $0<\varepsilon \ll 1$, then lubrication theory should be applicable and reasonably accurate in the gap region.
In this region, the surfaces of the two cylinders are at $x=\pm h(y)$, where
\be\label{boundaries}
h(y) = \varepsilon +\thalf y^2 +\tn {O} (y^4)\,,\quad dh/dy =y + \tn {O} (y^3)\,.
\ee
We shall use a hat $\hat{}$ throughout to denote kinematic equivalents of dynamic quantities; thus for example pressure 
$p=\mu \hat{p}$, force $F=\mu \hat{F}$, etc.   In the lubrication approximation (\cite{B67}, \S{4.8}), $\hat{p}=\hat{p}(y)$ and  $d\hat{p}/dy=\partial^2 v/\partial x^2$,
where ${\bf u} =(u,v)$ with $|v|\gg |u|$.
This integrates with boundary conditions $v =  \cos{y}\approx 1 $ on $x=\pm h(y)$ to give
\be\label{velprofile}
v(x,y)=1+ \thalf (d\hat{p}/dy)\left(x^2 -h(y)^2\right)\,. 
\ee
\vskip -2mm
\noindent The flux $Q$ between the cylinders is then
\vskip -5mm
\be\label{flux}
Q=\int_{-h(y)}^{h(y)}v(x,y)\,\tn{d}x =2\, h(y) - \frac{2}{3}\frac{d\hat{p}}{dy}\,h(y)^3\,,
\ee
\vskip -5mm
\noindent so that
\vskip -5mm
\be\label{pressuregradient}
 \frac{2}{3}\frac{d\hat{p}}{dy}=\frac{2}{h(y)^2}-\frac{Q}{h(y)^3} =\frac {2}{ (\varepsilon +\thalf y^2)^2}-\frac{Q}{ (\varepsilon +\thalf y^2)^3}\,.
\ee
This integrates to give the deviation of pressure from the pressure `at infinity' as
\be\label{pressure1}
\hat{p}(y)=-\frac{3Q y}{2 \varepsilon (y^2 + 2 \varepsilon)^2} - \frac{3 (3 Q- 8  \varepsilon)}{8 \varepsilon^2} \left[\frac{y}{y^2 + 2 \varepsilon}  +\frac{1}{ \sqrt{2 \varepsilon}}  \tan^{-1}{\left[y/\sqrt{2 \varepsilon}\right]}\right]\,.
\ee
For large $|y|$, (\ref{pressure1}) gives
\vskip -5mm
\be
\hat{p}(y)\sim\pm\left(3\pi/16\sqrt{2}\right)(3Q-8\varepsilon)\varepsilon^{-5/2}+\tn{O} (|y|^{-3})\,,
\ee
and since $\hat{p}(y)\rightarrow 0$ for large $|y|$, the leading term must vanish, giving $Q=8\,\varepsilon/3$.
As expected, $Q\rightarrow 0$ as $\varepsilon\rightarrow 0$.

\begin{figure}
\begin{center}
\includegraphics*[width=0.8\textwidth,  trim=0mm 0mm 0mm 0mm]{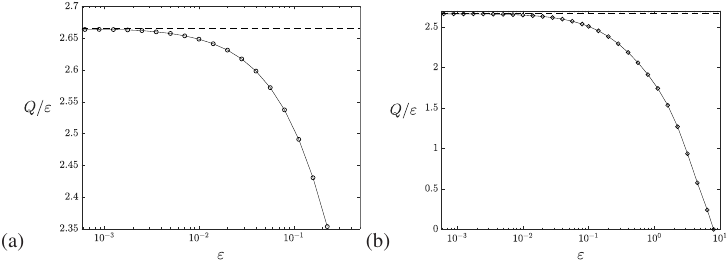} 
\end{center}
\caption{(a) Normalised mass flux $Q$ through the small gap between the cylinders as a
function of $\varepsilon$; the dashed asymptote is at $Q/\varepsilon=8/3 =
2.66\dots$, as determined by lubrication theory ; (b) showing that, as $\varepsilon$ \emph{increases} to its limiting value 
 (8 for $R_0=10$, when $\mathcal{C}_{1}$  and $\mathcal{C}_{2}$ make contact with  $\mathcal{C}_{0}$),  the normalised mass flux decreases to zero.}
\label{Flux_num}
\end{figure}
\begin{figure}
\begin{center}
\includegraphics*[width=0.8\textwidth,  trim=0mm 0mm 0mm 0mm]{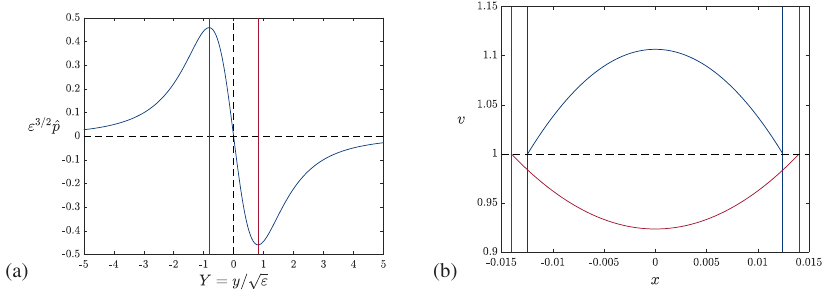}   
\end{center}
\caption{(a) Plot of $\varepsilon^{3/2}\hat{p}(Y)$ {\emph vs} $Y= y/\sqrt{\varepsilon}$ as given by (\ref{pressure3}); (b) velocity profiles (equation (\ref{velprofile2})) for $\varepsilon=10^{-2}$ at two sections, $y= 0.7 \varepsilon^{1/2}$  (blue) and $y=0.9\, \varepsilon^{1/2}$  (red); the boundaries are at $x=\pm h(y)$ in each case, and diverge as $y$ increases;  the viscous wall stress is positive for 
$y< 0.816\,\varepsilon^{1/2}$, negative for  $y>0.816 \, \varepsilon^{1/2}$.} 
\label{lubrication_plots}
\end{figure}
The flux $Q$ as a function of $\varepsilon$, computed from the  numerical
solution, is shown in figure \ref{Flux_num}(a,b) in which
the asymptote at $Q/\varepsilon \sim 8/3 \approx 2.666$ is shown by the
dashed line.  The behaviour here gives confidence that lubrication theory
is indeed reliable when $\varepsilon \ll 1$, and that  the flow in the gap
is well resolved by the numerical solution  for values of $\varepsilon$
down to a few times $10^{-5}$. For smaller values, numerical resolution becomes
more challenging, but the lubrication description becomes increasingly accurate.
\begin{figure}
\begin{center}
\includegraphics*[width=0.7\textwidth,  trim=0mm 0mm 0mm 0mm]{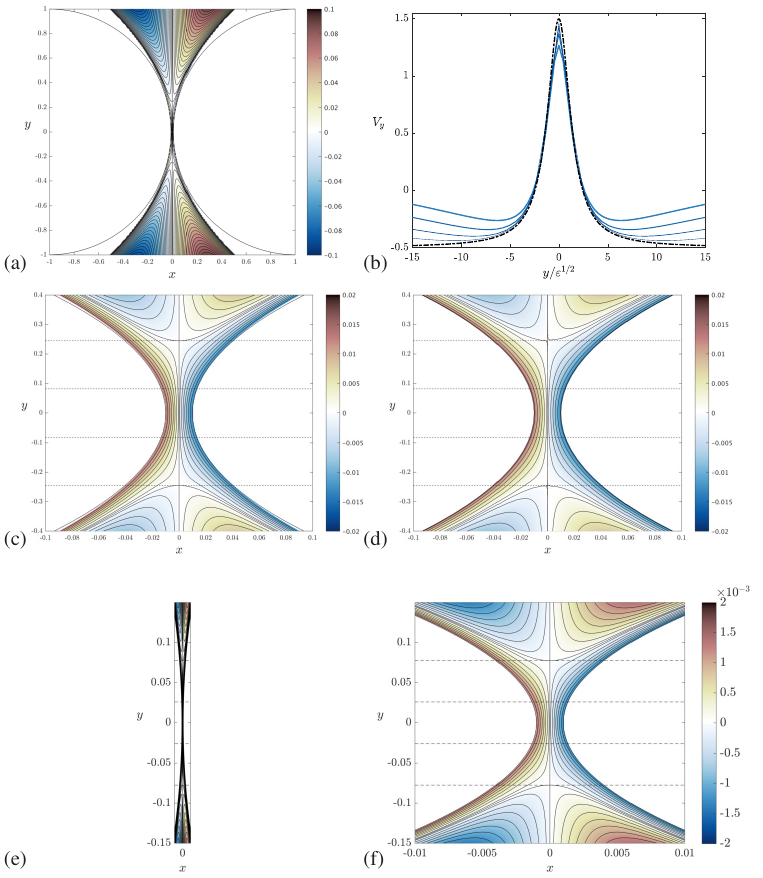}
\end{center}
\caption{(a) Streamlines $\psi_{L}(x,y)=$ const.~as given by (\ref{stream}) for
  the case $\varepsilon=0.01$ (compare with figure \ref{2pannels_num}b);
  (b) vertical velocity on the mid-plane $v(0,Y)$, where $Y= y /\sqrt{\varepsilon}$,  for 
  $\varepsilon = 0.2,\, 0.1,\, 0.05,\, 0.025,$ with correspondingly decreasing line thickness;
  the solution determined by lubrication theory (equation (\ref{vax})) is
 shown in black, dash-dotted;
  (c) same as (a), but closer to the origin and with $x$-coordinate stretched by a factor $10$;
  (d) the same from the full numerical solution, with the same colour code; the difference is almost imperceptible. The critical levels $y=\pm 0.816\,\varepsilon ^{1/2}$ and $y= \pm 2.45 \,\varepsilon ^{1/2}$ are shown by the dashed lines; 
  (e) and (f) the same for the case $\varepsilon=0.001$; 
  (e) correct aspect ratio for this value of $\varepsilon$; 
  (f) zoomed near the origin and stretched (by a factor 15) in the $x$-direction, to show more details of the structure; the critical levels $y=\pm 0.816 \,\varepsilon ^{1/2}$ and $y=\pm 2.45\, \varepsilon ^{1/2}$ are again shown by the dashed lines. } 
  \label{Streamlines_lub}
\end{figure}
With  $Q=8\varepsilon/3$,  (\ref{pressuregradient})  and (\ref{pressure1}) give
\vskip -5mm
\be\label{press_grad}
d\hat{p}/dy =[h(y)]^{-2}\left[3-4\varepsilon/h(y) \right]\,,
\ee
\vskip -5mm
\noindent and, with $Y= y/\sqrt{\varepsilon}$,
\vskip -5mm
\be\label{pressure3}
\hat{p}(y)=-\frac{4y}{(y^2+2\varepsilon)^2}\,,\quad \tn{or equivalently}\quad\varepsilon^{3/2}\,\hat{p}(Y)=-\frac{4Y}{(Y^2 +2)^2}\,,
\ee
\vskip -5mm
\noindent and (\ref{velprofile}) then gives
\vskip -5mm
\be\label{velprofile2}
v(x,y)=1+\frac{1}{2 h(y)^2}\left(3-\frac{4\varepsilon} {h(y)} \right)\left(x^2 - h(y)^2\right) \,.
\ee
Using the suffix $L$ to denote the lubrication approximation, the corresponding streamfunction $\psi_{L}(x,y)$, satisfying $\partial \psi_{L}/\partial x\!=\!-v(x,y)$  and $\psi_{L} (0,y)\!=\!0$, is
\be\label{stream}
\psi_{L}(x,y)=-x-\frac{1}{2 h(y)^2}\left(3-\frac{4\varepsilon} {h(y)} \right)\left(\tthird x^3 - h(y)^{2} x\right)\,.
\ee

The scaled pressure as a function of  $Y= y/\sqrt{\varepsilon}$ is shown in figure \ref{lubrication_plots}(a).  The singular behaviour
$p(y)\sim -4\,y^{-3}$ when $\varepsilon=0$ is unphysical, and indicates that in practice some deformation of the cylinders must occur if they are brought into contact while rotating, a phenomenon first recognised by Hertz (1882) \cite{H82}; moreover, the liquid will cavitate where $p(y)+p_{0}$ falls below the vapour pressure in the region $\{y>0, y\ll 1\}$.  The singularity in pressure could presumably be resolved by taking such effects into account.  This resolution problem lies outside the scope of the present treatment, and will be ignored in what follows.

The pressure gradient (\ref{press_grad}) is negative for $h(y)<4\,\varepsilon/3$,\, i.e.~for $|y|< (2\,\varepsilon/3)^{1/2} \approx 0.816\, \varepsilon ^{1/2}.$ It follows from (\ref{velprofile}) that the curvature of the velocity profile across the gap changes sign at $|y|\approx  0.816\, \varepsilon ^{1/2}$.  Figure~\ref{lubrication_plots}(b) shows two velocity profiles across the gap, the blue one at $y=0.7\, \varepsilon ^{1/2}$, just below the critical level at which the curvature changes sign, the red one at  $y=0.9\, \varepsilon ^{1/2}$ just above.  The wall stress below  the critical level $y=0.816 \,\varepsilon ^{1/2}$ evidently provides a positive contribution to the force on the cylinders, while above this critical level, it  provides a negative contribution.
The velocity on the axis $x=0$ is 
\be\label{vax}
v(0,y) =1+ \frac{1}{2}\left(\frac{4\varepsilon}{h(y)}-3\right)  = \frac{2  \varepsilon}{h(y)} -\frac{1}{2} = \frac{4}{2+Y^2} -\frac{1}{2}\,.
\ee
This vanishes where $h(y) =4\,\varepsilon$, i.e.~at $Y=\pm\sqrt{6}$, or $y=\pm (6\varepsilon)^{1/2}\approx \pm2.45 \,\varepsilon^{1/2}$; this  therefore, as previously stated, gives the location of the two saddle points on the $y$-axis.

Figure \ref{Streamlines_lub}(a) shows the streamlines $\psi_{L}$= const.~given by (\ref{stream}) for $\varepsilon=0.01$, which admits comparison with the numerical solution shown in figure \ref{2pannels_num}(b). Figure  \ref{Streamlines_lub}(c) shows the same nearer the origin  with the $x$-coordinate stretched by a factor 10, and figure \ref{Streamlines_lub}(d) shows the same streamlines as determined by the full numerical solution; the difference here is almost imperceptible, again giving confidence in the accuracy of both the numerics and the lubrication theory at least at this value of $\varepsilon=0.01$. In both panels, the critical levels $y=\pm 0.816\, \varepsilon ^{1/2}$ (at which the viscous wall stress changes sign) and $y = \pm 2.45\, \varepsilon ^{1/2}$ (at which the saddle points occur) are shown by the dashed  lines.

Figure \ref{Streamlines_lub}(b) shows the vertical velocity on the mid-plane $v(0,Y)$, where $Y= y /\sqrt{\varepsilon}$,  for 
 $\varepsilon = 0.2,\, 0.1,\, 0.05,\, 0.025$, with correspondingly decreasing line thickness, and the limiting curve (dash-dotted) as given by lubrication theory (equation (\ref{vax})). The curves shadow the limiting curve ever more faithfully as $\varepsilon$ decreases, giving further confidence in the relevance of the limiting lubrication-theory treatment.

Figure \ref{Streamlines_lub}(e) shows the streamlines in the gap as given by (\ref{stream}) for the even smaller value $\varepsilon =0.001$, and figure \ref{Streamlines_lub}(f) shows the same zoomed near the origin and stretched in the $x$-direction, again revealing the saddle-point structure.

\subsubsection{3.3 Force on the cylinders, as determined by lubrication theory}\label{force_cyl}
\noindent The vertical force $\mu F_y$ on each cylinder consists  of two parts: (i) the shear stress (viscous) force $ F_{yv}=\mu  \hat{F}_{yv}$ and (ii) the pressure force $ F_{yp}=\mu \hat{F}_{yp}$; for $\varepsilon > 0$, these are given by
\be\label{viscousforce}
\hat{F}_{yv} =-\int_{-\infty}^{\infty}\left.\frac{\partial v}{\partial x}\right|_{x=h(y)}\!\!\tn{d}y\,= \int_{-\infty}^{\infty}\frac{4\varepsilon-6y^2 }{(y^2+2\varepsilon)^2}\,\tn{d}y
=-\pi  (2/\varepsilon)^{1/2}\,,
\ee
and
\be\label{pressureforce}
\hat{F}_{yp} =\int_{-\infty}^{\infty}\hat{p}\left(-\frac{dh}{dy}\right)\,\tn{d}y =  \int_{-\infty}^{\infty}\frac{4y^2}{(y^2+2\varepsilon)^2}\,\tn{d}y
=\pi  (2/\varepsilon)^{1/2}\,,
\ee
so that
\be\label{totalforce}
\hat{F}_{yv}+\hat{F}_{yp} = \int_{-\infty}^{\infty}\frac{4\varepsilon-2y^2 }{(y^2+2\varepsilon)^2}\,\tn{d}y = 0\,.
\ee 
 
\begin{figure}
\begin{center}
\includegraphics*[width=0.7\textwidth,  trim=0mm 0mm 0mm 0mm]{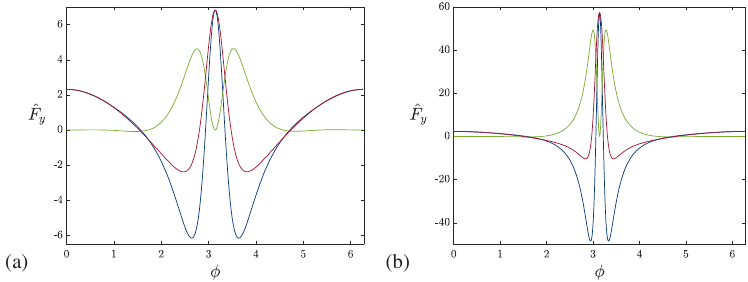} 
\end{center}
\caption{Distribution of the vertical forces  on $\mathcal{C}_{2}$ as functions of the angle $\phi$; the narrowest point of the gap is at $\phi=\pi$; $R_{0}=10$;  viscous term (blue),  pressure term  (green), total force (red);
  (a) $\varepsilon=10^{-1}$; (b) $\varepsilon=10^{-2}$. }
  \label{Fig_integrands}
\end{figure}
\begin{figure}
  \begin{center}
    \includegraphics*[width=\textwidth,  trim=0mm 0mm 0mm 0mm]{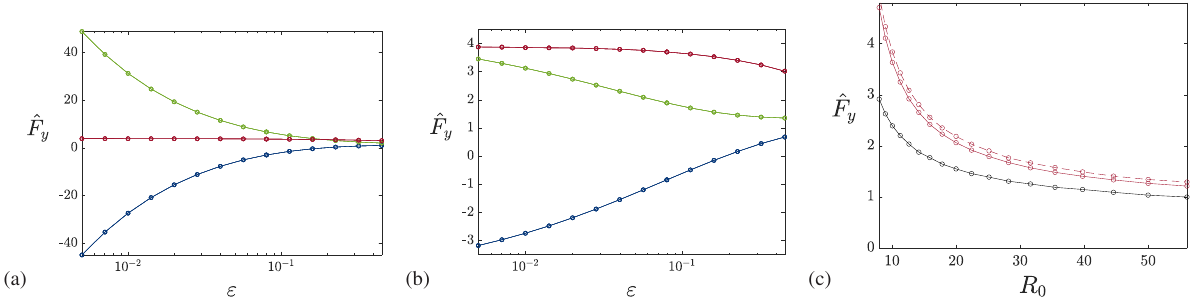}
\end{center}
\caption{(a) Integrated contributions to the
 vertical force on $\mathcal{C}_2$, viscous (blue), pressure (green) and the total (red), as  functions of
  $\varepsilon$; (b) the same, but with the viscous and pressure
  contributions rescaled by $\sqrt{\varepsilon}$ (the total force is not rescaled); (c) dependence of 
  total force on $R_0$,   $\varepsilon=0.1$  (solid), $\varepsilon=0.01$ (dashed);   no-slip condition on $\mathcal{C}_{0}$ (red),  stress-free condition on $\mathcal{C}_{0}$  (black).}
  \label{Vertical_force}
\end{figure}
Thus, somewhat surprisingly, the total force on each cylinder vanishes at  O$(\varepsilon^{-1/2})$. This conclusion is however supported by the numerical solution:  figure \ref{Fig_integrands} shows the distribution of the two contributions to the vertical force as functions of the angle $\phi$ round $\mathcal{C}_{2}$, confirming that the forces are indeed increasingly concentrated near the minimum gap position $\phi=\pi$ as $\varepsilon$ decreases; and figure \ref{Vertical_force} shows that when $R_0 =10$ the integrated pressure and viscous contributions do indeed nearly cancel at O$(\varepsilon^{-1/2})$, but that there is an O$(1)$ residual contribution $\hat{F}_{yv}+\hat{F}_{yp} \approx 3.9$ for $0<\varepsilon\ll 1$, a result that lies outside the scope of the  leading-order lubrication approximation.  We might expect that this  residual force should decrease to zero as $R_{0}$ increases without limit, in conformity with Jeffery's (1922) conclusion;  figure \ref{Vertical_force}(c) shows that the force does decrease at least in the range $8<R_{0}<55$;  this dependence implies that the force, being dependent on the far-field, cannot be obtained even by the higher-order, but still local, lubrication theory of Tavakol et al.~(2017) \cite{T2017}.  From comparison with the model problem of \S 2, it is almost certain that  $\hat{F}_{y}\sim (\log R_{0})^{-1}$ as $R_{0}$ increases, i.e. extremely slowly.  
{When $R_{0}\rightarrow\infty$ at fixed Reynolds number Re $\ll 1$,  inertia effects in the far field require that we replace $R_{0}$ by $r\sim \tn{Re}^{-1}\log{[\tn{Re}^{-1}}]$ [see (\ref{inertia_limit}) above]; it thus seems likely that the decrease of $\hat{F}_{y}$ will satisfy
\be
\hat{F}_{y}\sim  \left(\log \left[\tn{Re}^{-1}\log{[\tn{Re}^{-1}}]\right]\right)^{-1} \quad\tn{when}\,\, R_{0}\gg \tn{Re}^{-1}\gg1\,,
\ee 
although proof of this would require matching to an appropriate far-field solution.} 

\begin{figure}
  \centerline{\includegraphics*[width=0.45\textwidth,  trim=0mm 0mm 0mm 0mm]{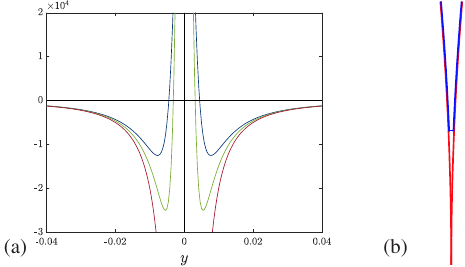}}
\caption{(a) Plot of the function $(4\varepsilon-2y^2)/(y^2+2\varepsilon)^2$, for $\varepsilon=10^{-5}$ (blue), $5\times 10^{-6}$ (green) and $\varepsilon= 0$ (red).  As $\varepsilon\downarrow 0$, the spike at $y=0$ becomes longer and narrower, and ultimately disappears in the limit $\varepsilon = 0$\,; (b) contour (shown in blue) used for calculation of the contact force when $\varepsilon=0$; the short segment is at the level $y=y_{1}$, and the limit $y_{1}\downarrow 0$ is considered.}
\label{Forceintegrand}
\end{figure}
But what if $\varepsilon=0$\,? \, In this situation, when the cylinders make contact,  the integrand in (\ref{totalforce}) is $-2/y^2$, and the integral  diverges.  The  force $\hat{F}_{yv}+\hat{F}_{yp}$ is then apparently infinite! The situation can be understood with reference to 
figure \ref{Forceintegrand}, which shows a plot of the integrand in (\ref{totalforce}) for $\varepsilon= 10^{-5}$ (blue), $5\times 10^{-6}$ (green) and $\varepsilon= 0$ (red).  As  $\varepsilon$ decreases the curves approach the limit curve more and more closely.   The spike in the region 
$|y| < (2 \varepsilon)^{1/2}$ contributes the positive value to the integral that exactly compensates the negative value from the  region $|y| > (2 \varepsilon)^{1/2}$.  When $\varepsilon=0$, the spike disappears, and only the negative contribution survives. 

However, in this limit situation when the cylinders make contact, there is
a pressure discontinuity across the point of contact, and this can
contribute a  force $\hat{F}_c$  to the resultant total force on the two
cylinders.
We describe this as a `flow-induced contact force', because it arises as a \emph {net} force \emph {on the pair of cylinders} due to the contact between them in the presence of flow. [This must be distinguished from the more familiar (equal and opposite) `solid contact force' that may be experienced by \emph {each} cylinder by virtue of their contact.] This flow-induced contact force can be obtained as follows.
\subsubsection{3.4 Flow-induced contact force}\label{Sec_Contact_force}
\noindent We need simply consider the restricted domain of fluid inside the blue contour shown in figure \ref{Forceintegrand}(b) for the contact situation $\varepsilon=0$ with $h(y)=\thalf y^2$.  The short segment in this figure is at the level $y=y_{1}$. The  total force on the curved parts of the blue contour is
\be\label{down_force}
2 \int_{y_{1}}^{\infty}\frac{(-2y^{2}) }{y^4}\,\tn{d}y = -4/y_{1}\,.
\ee
From (\ref{pressure3}), when $\varepsilon=0$, the pressure is given by $\hat{p}=-4/y^3 $, so the pressure force on the small horizontal segment (an upwards suction)  is
\be
\int_{-h(y_1)}^{h(y_1)}\!\!-\hat{p}\,dx =(4/y_{1}^3) y_{1}^2 = +4/y_{1}.
\ee
This exactly balances the force (\ref{down_force}) from the curved parts of the contour.  (These forces are $\pm 8/y_{1}$ when contributions from the region $y<0$ are  also taken into account.)  This force balance persists in the limit $y_{1}\downarrow 0$, and we may conclude that what was the upward  contribution to $\hat{F}_{yv}+\hat{F}_{yp} $ when $\varepsilon>0$ is replaced when $\varepsilon=0$ by the upward contact force resulting from the (infinite) jump in pressure across the point of contact.  [This argument ignores cavitation, which must break the symmetry of the limiting flow across $y=0$.]

The analytic treatment of this limit that now follows  provides an alternative derivation of this contact force that does not rely on the lubrication approximation.

\begin{figure}
\begin{center}
(a)\includegraphics[width=7cm, trim=0mm 0mm 0mm 0mm, clip]{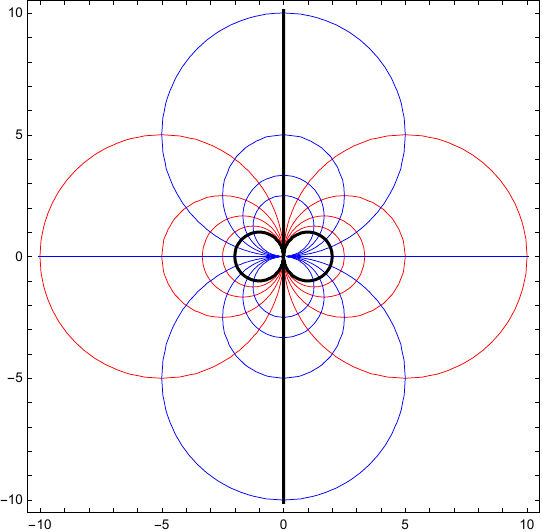}\quad
(b)\includegraphics[width=7cm, trim=0mm 0mm 0mm 0mm, clip]{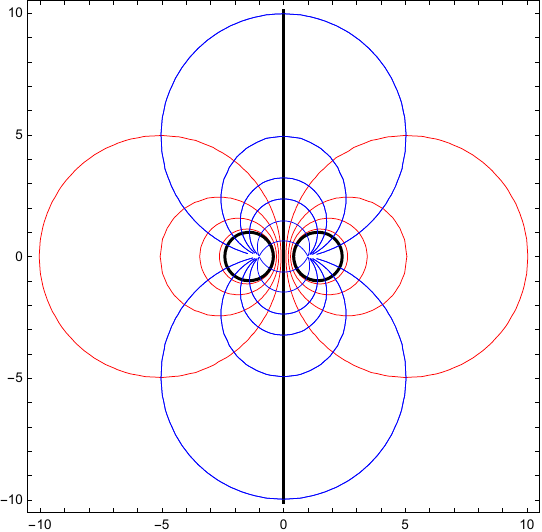}
\end{center}
\caption{(a) Contours $\xi=$ const. (red) and $\eta=$ const. (blue), given by (\ref{conformal_map}); the contours  $\xi=0,\,\pm 1/2$ are shown in black; (b) corresponding contours given by the conformal mapping 
$\zeta = \log [(z+c)/(z-c)]$, as used by Jeffery (1922); here $c=1$ and the cylinders of unit radius are shown by the contours  $\xi=\pm \sinh^{-1}{c} \approx \pm 0.881$.}
\label{contours}
\end{figure}

\vskip 5mm \noindent
{\bf 4 Analytic solution when $\varepsilon=0$ and $R_{0}=\infty$}
\vskip 2mm \noindent
When $\varepsilon=0$ and $R_{0}=\infty$, the Stokes problem may be solved exactly; this is the limit version  of Jeffery's (1922) problem (see Appendix A).  Following Schubert (1967) \cite{S67}, we use the 
 conformal mapping $\zeta\equiv \xi+\textnormal{i}\eta =1/z$, where $z\equiv x+\textnormal{i}y$, giving
\be\label{conformal_map}
\quad \xi=\frac{x}{x^2 +y^2}\,,\quad \eta=\frac{-y}{x^2 +y^2}\,.
\ee
The scale factor for this mapping is
\be
h(\xi,\eta)= |d\zeta/dz|=(x^2 +y^2)^{-1}=\xi^2 +\eta^2.
\ee
The contours $\xi=$ const., $\eta=$ const., are the circles shown in figure \ref{contours}(a). 
The essential property of this mapping is that  $\psi(x,y)$ satisfies the biharmonic equation $\nabla^4 \psi=0$ if and only if 
$\Psi(\xi,\eta)=h(\xi,\eta)\psi\left[x(\xi,\eta),y(\xi,\eta)\right]$ satisfies $\nabla_{\xi,\eta}^{4} \Psi=0$.

\begin{figure}
  \centerline{\includegraphics*[width=\textwidth,  trim=0mm 0mm 0mm 0mm]{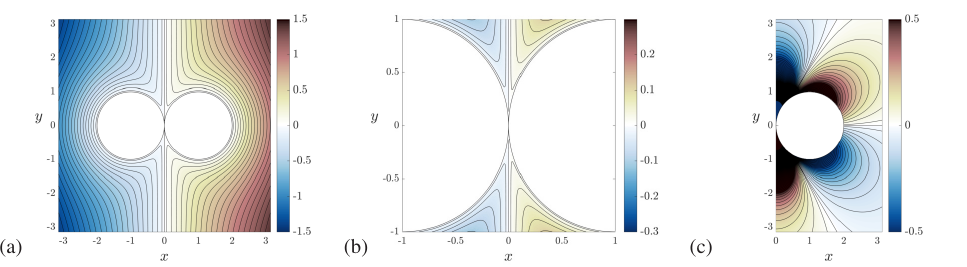}}
\caption{(a) Streamlines $\psi=$ const., as given by (\ref{streamfn_counter}); (b) zoom near the point of contact; (c) contours $ \hat{p}=$ const. in the half-plane $x>0$ as given by (\ref{pressurefield}). }
\label{contours2}
\end{figure}

For our problem, we want a solution of $\nabla^4 \psi=0$ that is antisymmetric about $x=0$, or equivalently a solution of 
$\nabla_{\xi,\eta}^{4} \Psi=0$ antisymmetric about $\xi=0$. We also need to impose angular velocities $\pm 1$ on the circles 
$\xi=\mp 1/2$.  The boundary conditions are then
\be
\Psi =\Psi_{\xi\xi}=0\quad \tn{on}\quad \xi=0,\quad\tn{and}\quad \Psi=0,\,\, \Psi_{\xi}= -1\quad  \tn{on}\quad \xi=\pm 1/2.
\ee
The required solution evidently does not depend on $\eta$; it is given by $\Psi(\xi,\eta)=\xi/2-2\xi^3$, and correspondingly
\be\label{streamfn_counter}
\psi(x,y)=(x^2 +y^2)\Psi(\xi,\eta)=x/2-2 x^{3}(x^2 +y^2)^{-2}=x/2- \frac{\partial}{\partial x}\left[\frac{y^{2}}{x^{2}+y^{2}}+\log{\left(x^{2}+y^{2}\right)}\right]\,.
\ee
In plane polar coordinates $\{r,\theta\}$, this is equivalently
\be\label{psi_polars}
\psi(r,\theta)=\thalf r\cos\theta -\thalf r^{-1}\left(3\cos\theta + \cos 3\theta\right)\,,
\ee
which does indeed satisfy  $\nabla^4 \psi=0$.   The streamlines $\psi=$ const. are shown in figure \ref{contours2}(a), and a zoom near the point of contact in figure \ref{contours2}(b).  The leading term $x/2$ in (\ref{streamfn_counter}) represents the Jeffery uniform stream [$(0,-\thalf)$ in this contact limit], while the term $-2 x^{3}(x^2 +y^2)^{-2}$ admits interpretation as the flow due to a `torque doublet' (or, to coin a suitable word comparable to stresslet \cite{LM16}, a `torquelet'): 
\be\label{torquelet}
\psi_{\mathcal{T}}=-(\partial/ \partial x) (\log {r^{2}} +\sin ^{2}{\theta})\,.
\ee 
The streamlines of the torquelet $\psi_{\mathcal{T}}=$ cst.~are shown in figure \ref{torquelet}(a); the {apparent compression of the streamlines towards the plane $y=0$} as compared with the streamlines of the simpler vortex dipole $\psi_{\mathcal{V}}= -(\partial/ \partial x) (\log {r}^{2})$ [figure \ref{torquelet}(b)] is evident. This {compression is due to the no-slip condition on the two rotating cylinders and their resulting mutual influence, an effect that evidently persists to the far field in the fluid.}
\begin{figure}
\begin{center}
(a)\includegraphics*[width=.4\textwidth, trim=10mm 4mm 0mm 0mm, clip]{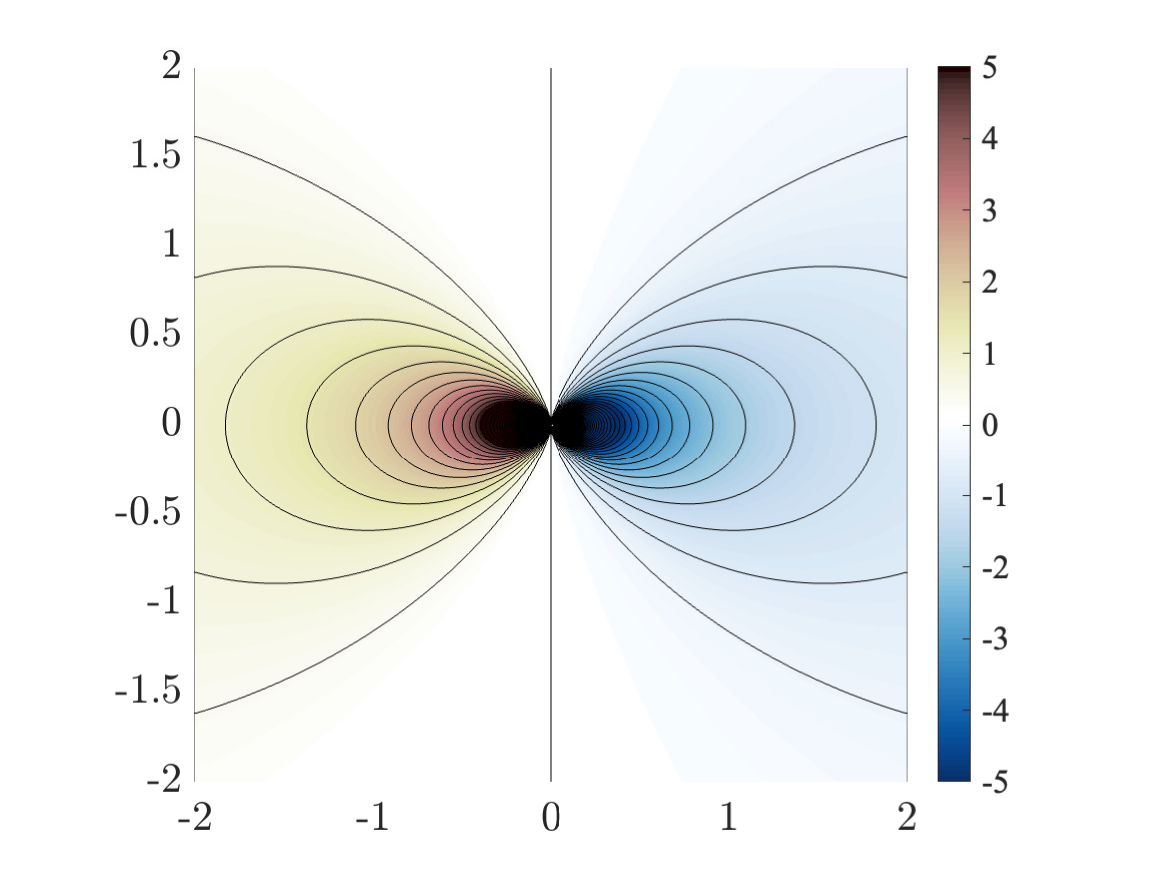}\qquad
(b)\includegraphics*[width=.4\textwidth, trim=10mm 4mm 0mm 0mm, clip]{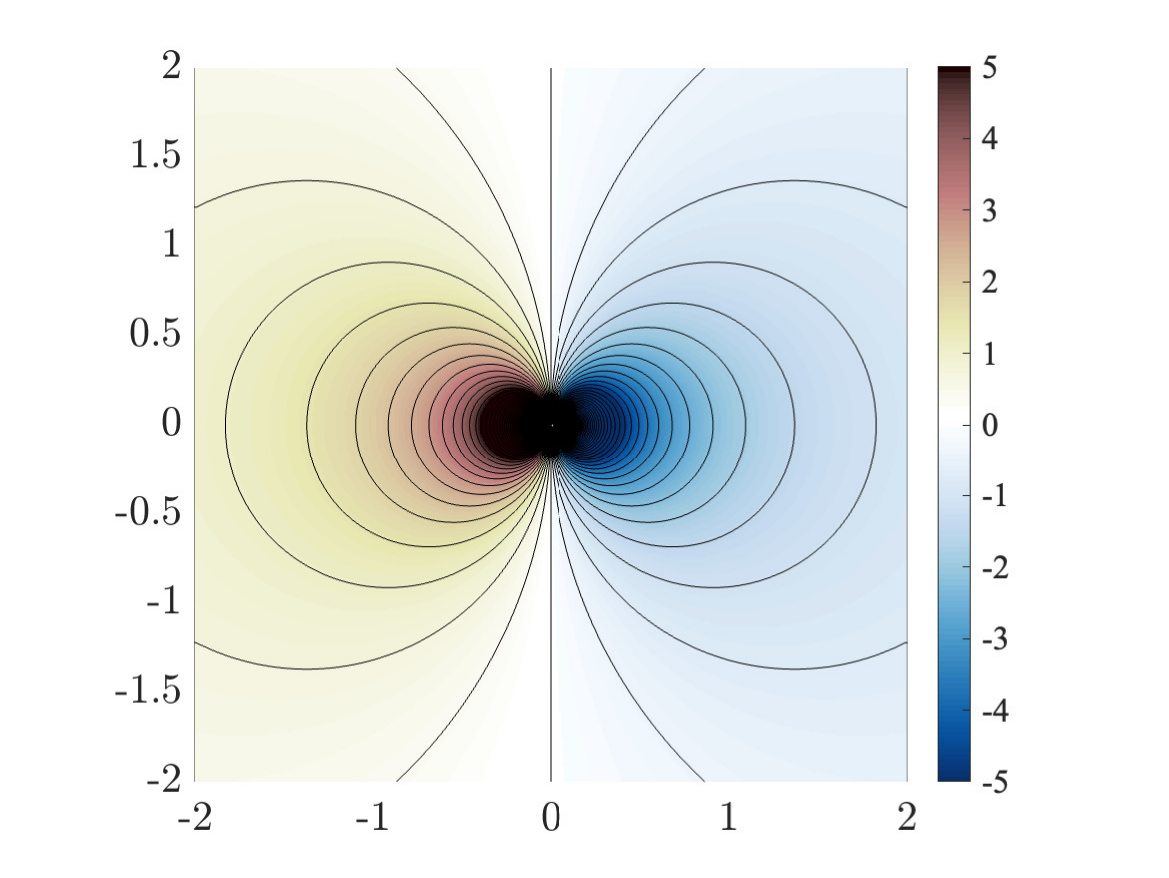}
\end{center}
\caption{(a) Streamlines associated with the torquelet $\psi_{\mathcal{T}}=-(\partial/ \partial x) (\log {r^{2}} +\sin ^{2}{\theta})$; (b) streamlines associated with the vortex dipole  $\psi_{\mathcal{V}}=-(\partial/ \partial x) (\log {r^{2}})$. {The torquelet streamlines appear to be more compressed towards the plane $y=0$.}}
\label{torqueletf}
\end{figure}
\subsubsection{4.1 Pressure field}
\noindent The velocity components are
\be \label{vel_comp}
u(x,y)=\frac{\partial \psi}{\partial y}=\frac{8x^3 y}{(x^2 +y^2)^3}\,,\quad
 v(x,y)=-\frac{\partial \psi}{\partial x}=-\frac{1}{2}-\frac{2(x^4 -3x^2 y^2)}{(x^2 +y^2)^3}\,,
\ee
and the pressure field satisfies the equations
\be
\frac{\partial \hat{p}}{\partial x} = \nabla^2 u=-\frac{48 x y (x^2 - y^2)}{(x^2 + y^2)^4}\,,\quad \frac{\partial \hat{p}}{\partial y} = \nabla^2 v =\frac{12(x^4 - 6x^2 y^2 +y^4)}{(x^2 + y^2)^4}. 
\ee
Either of these equations may be integrated, giving
\be\label{pressurefield}
 \hat{p}(x,y)=4 y ( 3 x^2 - y^2)(x^2 + y^2)^{-3}\,.
\ee
The contours  $\hat{p}(x,y)=$ const.~in the half-plane $x\ge 0$ are shown in figure \ref{contours2}(c).

On the right-hand cylinder  $\mathcal{C}_{2}$, let 
\be\label{cyl_phi}
x=1+\cos\phi,\,\,\,y=\sin\phi\,, \,\,(-\pi\le\phi\le \pi), \quad\tn{so}\,\,\, r^{2}\equiv x^2 +y^2 = 2(1+\cos\phi)\,.
\ee
The normal and tangent vectors on $\mathcal{C}_{2}$ have cartesian components
\be
{\bf n} =(\cos\phi,\sin\phi),\quad {\bf t} = (-\sin\phi, \cos\phi)\,.
\ee
From (\ref{pressurefield}), the pressure on $\mathcal{C}_{2}$ is then
\be \label{pressure_C2}
 \hat{p}(\phi) =\thalf \sin\phi\left[3(1+\cos\phi)^{2}-\sin^2\phi)\right](1+\cos\phi)^{-3} \,.
\ee
The vertical component of the pressure force  $- \hat{p}\, {\bf n}$ on $\mathcal{C}_{2}$ is $- \hat{p}(\phi)\sin\phi$, which has the asymptotic  behaviour
\be
- \hat{p}(\phi)\sin\phi \sim 4(\phi -\pi)^{-2}  \quad \tn{near} \quad \phi =\pi,
\ee
obviously non-integrable at the point of contact $\phi=\pi$.

\subsubsection{4.2 Viscous stress on cylinder}
\noindent The rate-of strain components are
\be
e_{11} =\partial u/\partial x,\quad e_{12} =e_{21}=\thalf(\partial u/\partial y+\partial v/\partial x),\quad e_{22} =\partial v/\partial y\,. 
\ee
Substituting (\ref{vel_comp}), and simplifying leads to the results
\be
e_{11}(x,y)=-e_{22}(x,y)= \frac{24y(x^2 y^2 - x^4)}{(x^2 +y^2)^4}\,,\quad e_{12}(x,y)=\frac{6x(x^4 -6 x^2 y^2 +y^4)}{(x^2 +y^2)^4}\,.
\ee
On the cylinder $\mathcal{C}_2$ with parametric equations (\ref{cyl_phi}), these reduce to
\be
e_{11}(\phi) =-e_{22}(\phi) =\frac{-3\sin{2\phi}}{2(1+\cos\phi)}\,,\quad e_{12}(\phi) =\frac{3\cos{2\phi}}{2(1+\cos\phi)}\,.
\ee
The viscous stress components $\tau_{1}(\phi)=\mu\,\hat{\tau}_{1}(\phi)$ and $\tau_{2}(\phi)=\mu\,\hat{\tau}_{2}(\phi)$   on $\mathcal{C}_2$  are given by
\be
\hat{\tau}_{1}(\phi)=2( e_{11}(\phi) \cos\phi +e_{12}(\phi)\sin\phi)\,,\quad \hat{\tau}_{2}(\phi)= 2(e_{21}(\phi) \cos\phi +e_{22}(\phi)\sin\phi)\,,
\ee
and these reduce to
\be\label{stress_cpts}
\hat{\tau}_{1}(\phi)=-3\sin\phi(1+\cos\phi)^{-1}\,,\quad \hat{\tau}_{2}(\phi)= {3\cos\phi(1+\cos\phi})^{-1}\,,
\ee
 with the asymptotic behaviour near $\phi=\pi$
\be\label{asymptotic_stress}
\hat{\tau}_{1}(\phi)\sim 6(\pi-\phi)^{-1}\,,\quad \hat{\tau}_{2}(\phi)\sim-6(\pi-\phi)^{-2}\,.
\ee
Note that the total vertical component of stress on $\mathcal{C}_2$, $-\hat{p}(\phi)\sin\phi +\hat{\tau}_{2}(\phi)$, therefore behaves like $-2(\pi-\phi)^{-2}$ near $\phi-\pi$, and is therefore non-integrable, indicating an infinite integrated downward contribution to the vertical force.

\subsubsection{4.3 Confirmation of flow-induced contact force}
\begin{figure}
\begin{center}
(a)\includegraphics*[height=6cm, trim=0mm 0mm 0mm 0mm, clip]{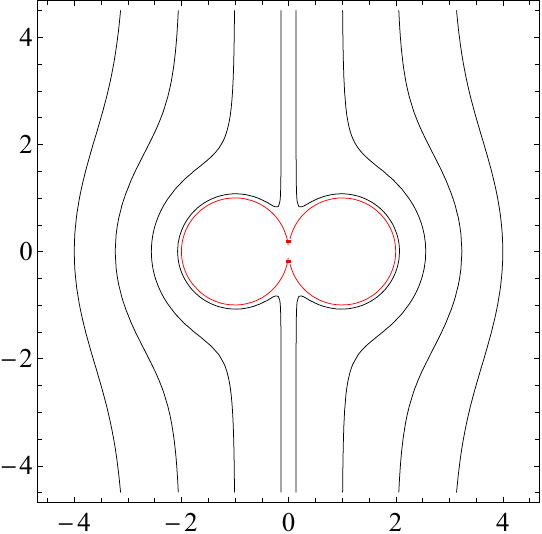} \qquad\quad
(b)\includegraphics*[height=6cm, trim=0mm 0mm 0mm 0mm, clip]{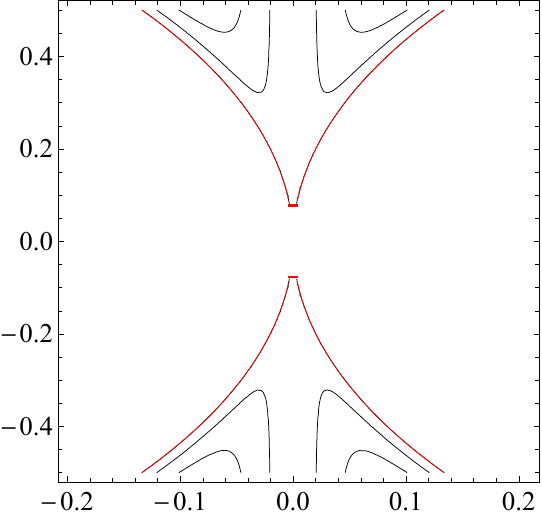} 
\end{center}
\caption{(a) Contour of integration (shown in red) for calculating the total resultant vertical force on both cylinders when they make contact ($\varepsilon=0$); the integral round the right-hand cylinder $\mathcal{C}_2$ runs from $\phi=-\phi_1$ to $\phi=\phi_1$, where $0<\pi-\phi_1\ll 1$ (and this is equal, by symmetry, to the contribution from $\mathcal{C}_1$); and the integral on the small horizontal segments at $y=\pm y_1$ run from $x=-y_{1}^{2}/2$ to $y_{1}^{2}/2$, where $y_1=\sin{\phi_1}=\sin{(\pi-\phi_1)} \sim \pi-\phi_1$; we then take the limit $\phi_{1}\rightarrow\pi$ (i.e. $y_{1}\rightarrow 0$); (b) zoom near the origin, and expanded by a factor of 2 in the $x$-direction to make the short sections at $y=\pm y_1$ more clearly visible.}
\label{Integration_contour}
\end{figure}

\noindent In order to resolve this singularity, we  consider the vertical force now integrated round the closed contour shown in red in
 figure \ref{Integration_contour}, as described in the figure caption.
The integral of the vertical force component round $\mathcal{C}_2$ from $\phi=-\phi_1$ to $\phi=\phi_1$ is, with some simplification,
\be
\int_{-\phi_1}^{\phi_1}(-\hat{p}(\phi)\sin\phi +\hat{\tau}_{2}(\phi)) \,\tn{d}\phi = \int_{-\phi_1}^{\phi_1}\frac{2\cos{\phi}+\cos{2\phi}}
{1+\cos{\phi}}\,\tn{d}\phi = 2 \sin (3\phi_{1}/2)\sec(\phi_{1}/2)\,.
\ee
  The contribution to $\hat{F}_{yp}+\hat{F}_{yv}$ from both $\mathcal{C}_1$ and $\mathcal{C}_2$ is therefore
\be\label{C1C2}
4 \sin{(3\phi_{1}/2)}\sec{(\phi_{1}/2)} \sim  -8(\pi-\phi_1)^{-1} +\twentysixthirds(\pi-\phi_1) +\tn{O}(\pi-\phi_1)^3\,,
\ee
which is indeed singular as $\phi_{1}\rightarrow \pi$.   However, as in \S 3.2, there is also a contribution to the total vertical force from 
 the small segments at $y=\pm y_1$; this is
\be\label{contact_force}
\hat{F}_{c} =\int_{-h(y_1)}^{h(y_1)}[\hat{p}(x,-y_{1})-\hat{p}(x,y_1)]\,\tn{d}x = \frac{128}{y_{1}(4+y_{1})^2} \sim \frac{8}{y_1} -4y_{1}+\tn{O}(y_{1}^3)\,, 
\ee
and, since $y_{1}\sim \pi-\phi_{1}$, the singularity here exactly cancels the singularity in (\ref{C1C2}) as $y_{1}\rightarrow 0$.  Moreover since this leaves terms of order $y_{1}$ which vanish in the limit, it follows that, in the limit, the total net force on the composite body is zero:
\be
\hat{F}_{y}=\lim_{y_1\rightarrow 0}(\hat{F}_{yp} +\hat{F}_{yv} +\hat{F}_{c}) =0\,.
\ee
This is consistent with the result (\ref{totalforce}), implying that the conclusion that the force is zero is in fact valid for all $\varepsilon\ge 0$.  This  also confirms the validity of the previous simpler derivation of the same force balance under  the lubrication approximation. We shall see in \S5 below how $\thalf \hat{F}_{c}$ exerts part of the torque that each cylinder experiences about its axis.

Note that in a frame of reference fixed in the fluid at infinity, the two cylinders move with velocity ${\bf V}=(0,\thalf)$.  The force generated by their rotation is then equal and opposite to the drag force that they jointly experience as they move through the fluid; in effect the velocity ${\bf V}$ is just such that this force balance is exactly satisfied.

\vskip 5mm \noindent
{\bf 5 Torque on the cylinders}
\subsubsection{5.1 Torque when $\varepsilon>0$} \label{torque_epsilon_pos}
\noindent Since the total force on each cylinder, including the contact force, is zero (when $R_{0}\rightarrow\infty$), the torques $ \hat{\mathcal{T}}_{1}\,{\bf e}_z$ and 
$\hat{\mathcal{T}}_{2}\,{\bf e}_z=\hat{\mathcal{T}}_{1}\,{\bf e}_z$ acting on $\mathcal{C}_{1,2}$ are independent of the point relative to which the torque is calculated, e.g.  
$\hat{\mathcal{T}}_{2}$  is the same whether calculated relative to the centre of $\mathcal{C}_2$ or relative to the origin O. When  $0<\varepsilon \ll 1$, the viscous drag on $\mathcal{C}_{2}$ is concentrated near the gap point $(\varepsilon, 0)$  and  the pressure on $\mathcal{C}_{2}$ makes no contribution to the torque about its axis at $(1+\varepsilon,0)$; hence,  using (\ref{viscousforce}), this torque is given  by 
\be\label{torque_asymp}
\hat{{\mathcal T}}_{2} \sim -\hat{F}_{yv} \sim \pi  (2/\varepsilon)^{1/2}\,.
\ee 
\begin{figure}
\begin{center}
\includegraphics*[width=0.7\textwidth,  trim=0mm 0mm 0mm 0mm]{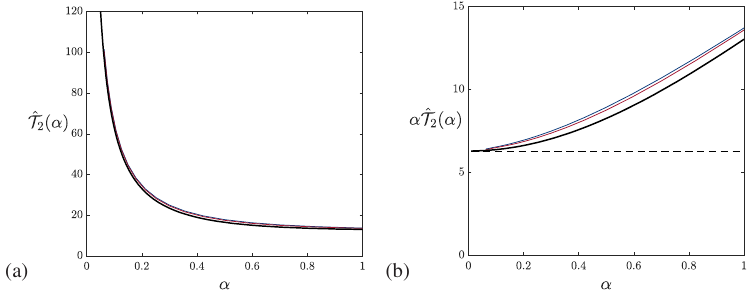} 
\end{center}
\caption{(a) The torque $\hat{\mathcal{T}}_{2}(\alpha)$ where $\alpha=\cosh^{-1}(1+\varepsilon)\sim (2\varepsilon)^{1/2}$, computed
from Jeffery's solution (see Appendix A, in black), and from the full numerics with $R_{0}=10$ (in blue), and
$R_{0}=15$ (in red); (b) The function $\alpha \hat{\mathcal{T}}_{2}(\alpha)$ with the same colour code; the level $2\pi$ is shown by the dashed line, showing that   
$\hat{\mathcal{T}}_{2}(\alpha)\sim 2\pi/\alpha = \pi(2/\varepsilon)^{1/2}$ as $\alpha\rightarrow 0$.} 
 \label{Torque_J}
\end{figure}
\!\!Figure \ref{Torque_J}(a) shows the torque $\hat{\mathcal{T}}_{2}(\alpha)$ (where $\alpha\sim (2\varepsilon)^{1/2}$) computed both from our  numerical solution for $R_{0}=10$ (blue) and $15$ (red), and from Jeffery's solution \cite{J22} as described in 
Appendix A (black), and figure \ref{Torque_J}(b) shows the corresponding compensated functions  $\alpha\,\hat{\mathcal{T}}_{2}(\alpha)$. The convergence as $R_{0}$ increases is evident, and  it is clear  that, in the limit $R_{0}\rightarrow\infty$,   $\hat{\mathcal{T}}_{2}(\alpha)\sim 2\pi/\alpha$ as $\alpha\rightarrow 0$, in perfect agreement with the asymptotic result (\ref{torque_asymp}). 

The pair of torques $\pm \hat{\mathcal{T}}_{2}(\alpha)$ do indeed constitute a torquelet associated with the term 
$-\thalf r^{-1}\left(3\cos\theta + \cos 3\theta\right)$ in (\ref{psi_polars}). 
 We shall confirm in Appendix A that (\ref{psi_polars}) is the limit as $\varepsilon\rightarrow 0$ of the solution found by Jeffery (1922) and that the torquelet ingredient of this flow is associated in the limit  $\alpha\sim(2\varepsilon)^{1/2}\rightarrow 0$ with torque singularities $\pm \hat{\mathcal{T}}_{2}(\alpha)\sim\pm 2\pi/\alpha$ separated by the vanishing distance $d(\alpha)\sim3\,\alpha$.

\subsubsection{5.2 Torque when $\varepsilon=0$}\label{Sec_torque_epsilon_zero}
\noindent When $\varepsilon=0$, it is instructive also to calculate the asymptotic behaviour of this torque as a function of the cut-off level $y_1$ defined in figure  \ref{Integration_contour}.  From the cartesian stress components (\ref{stress_cpts}), the tangential stress on $\mathcal{C}_{2}$ is 
\be
\hat{\tau}(\phi) = \hat{\tau}_{2}(\phi) \cos\phi - \hat{\tau}_{1}(\phi) \sin\phi = 3(1+\cos\phi)^{-1}\,.
\ee
Integrating this from $-\phi_1$ to $+\phi_1$ where $\pi-\phi_{1}\sim y_{1}\ll 1$ gives the moment of this tangential stress as
\be
\hat{M}_{v}=\int_{-\phi_{1}}^{\phi_{1}} \frac{3}{1+\cos\phi}\,\tn{d}\phi = 6\tan (\phi_{1}/2)\sim \frac{12}{\pi-\phi_{1}}\quad \tn{as}\,\,\phi_{1}\rightarrow\pi\,.
\ee
There is also a contribution from the moment of the half of the contact force $\hat{F}_{c}$ that can be deemed to act on $\mathcal{C}_{2}$, viz,
$\hat{M}_{c}= -\thalf \hat{F}_{c} \sim -4/y_{1}$ from (\ref{contact_force}). Combining these moments gives the torque
\be\label{torque_O1}
\hat{\mathcal{T}}_{2}(y_1) = \hat{M}_{v}+\hat{M}_{c} \sim 8/y_{1}\quad \tn{as}\,\,y_{1}\rightarrow 0\,,
\ee
and we note again that, by symmetry, $\hat{\mathcal{T}}_{1}(y_1)=-\hat{\mathcal{T}}_{2}(y_1)$.

Alternatively, we may calculate this same torque relative to the origin O.  The tangential stress $\hat{\tau}(\phi)$ exerts a moment $(\bf{x}\times \bf{t}) \hat{\tau}(\phi)$ about the contact point O$(0,0)$;  here,
 \be
 {\bf x}=(1+\cos\phi,\sin\phi)\,\,\,\,\tn{and}\,\,\,\,  {\bf t}=(-\sin\phi,\cos\phi)\,,\quad\tn{so} \quad{\bf x}\times {\bf t} =(1+\cos\phi) {\bf e} _{z}\,,
 \ee
 and it follows remarkably that this moment takes the value $3$, uniform on $\mathcal{C}_{2}$, so that the total moment of the tangential stress about O takes the finite value $\hat{M}_{\tn{O}v}=3(2\pi)=6\pi$.  However, noting that
 \be \label{x_cross_n}
{\bf x}\times {\bf n}= (1+\cos{\phi},\sin{\phi})\times (\cos{\phi},\sin{\phi})=\sin{\phi}\,\,{\bf e}_{z} \quad\tn{on}\,\,\mathcal{C}_{2}\,,
\ee
  it is actually the pressure $-\hat{p}(\phi)\bf{n}$ on $\mathcal{C}_{2}$ that exerts the dominant moment about O; using (\ref{pressure_C2}) and (\ref{x_cross_n}), this is 
 \be
 \hat{ M}_{\tn{O}p}=\int_{-\phi_{1}}^{\phi_{1}}  -\hat{p}(\phi) \sin\phi \,\tn{d}\phi = -6\phi +4 \sin\phi +4\tan(\phi/2)\sim \frac{8}{\pi-\phi_{1}} -6\pi \,,
\ee
 and so the torque relative to O is
 \be\label{torque_asymp_2}
\hat{ \mathcal{T}}_{2}(y_1) = \hat{M}_{\tn{O}v}+ \hat{ M}_{\tn{O}p} \sim 8/y_{1} \quad\tn{as}\,\,y_{1}\sim\pi-\phi_{1}\rightarrow 0\,,
 \ee 
in precise agreement with (\ref{torque_O1}).  We should note here that the
flow-induced contact force has zero moment about O in the limit $y_{1}\rightarrow 0$. 

\section{Part II: Co-rotating cylinders}
\vskip 2mm \noindent
{\bf 6. An elementary model problem}
\vskip 2mm \noindent
Figure 15(a) shows the symmetries of the flow induced by co-rotating cylinders.
We recall first an elementary model problem that exhibits the same symmetries as those of this figure, namely the problem of flow in an annulus $1<r<R_{0},\, R_{0}\gg 1$, when the inner cylinder $r=1$ rotates with angular velocity $\Omega=-1$, and the outer cylinder $r=R_{0}$ is at rest. The flow in the annulus is  purely azimuthal, 
\be
\bfu=[0, v(r)] \quad\tn{where} \,\,v(r)= Ar +B/r \quad \tn{with} \,\,A+B=-1\,.
\ee
There are two quite distinct possible scenarios:
\vskip 1mm
\noindent (i) We may impose a `zero-stress' condition, $\tn{d}(v/r)/\tn{d}r=0$, on $r=R_{0}$; then $B=0$ and $v(r)=-r$, rigid-body rotation throughout the fluid, obviously non-zero at infinity when $R_{0}\rightarrow\infty$. The torque acting on the cylinder is zero in this situation. This is the scenario that is apparent in Jeffery's (1922) approach to the two-cylinder problem described in \S \S \,8.1 and  8.2 below.
\vskip 1mm
\noindent (ii) Alternatively, we may impose no-slip on $r=R_{0}$; then  $AR_{0} +B/R_{0}=0$, giving $ v(r)=(r-R_{0}^{2}/r)/(R_{0}^{2}-1)$. Thus, for $1\le r\ll R_{0}$,  we have $v(r)\sim -1/r$, the flow due to a virtual point vortex at $r=0$. In the limit $R_{0}\rightarrow\infty$, this point-vortex flow extends throughout the fluid tending to zero at infinity, and the fluid exerts a nonzero torque $\mathcal {T}_{s}=4\pi\mu[1+\tn{O}(R_{0}^{-2})]$ on the cylinder. This is the scenario that is evident in Watson's (1995) remedy (\S 8.3 below), which incorporates the appropriate torque that nullifies the far-field rigid-body rotation.

\begin{figure}
\begin{center}
(a)\includegraphics*[height=5cm,trim=14mm 6mm 15mm 0mm,clip]{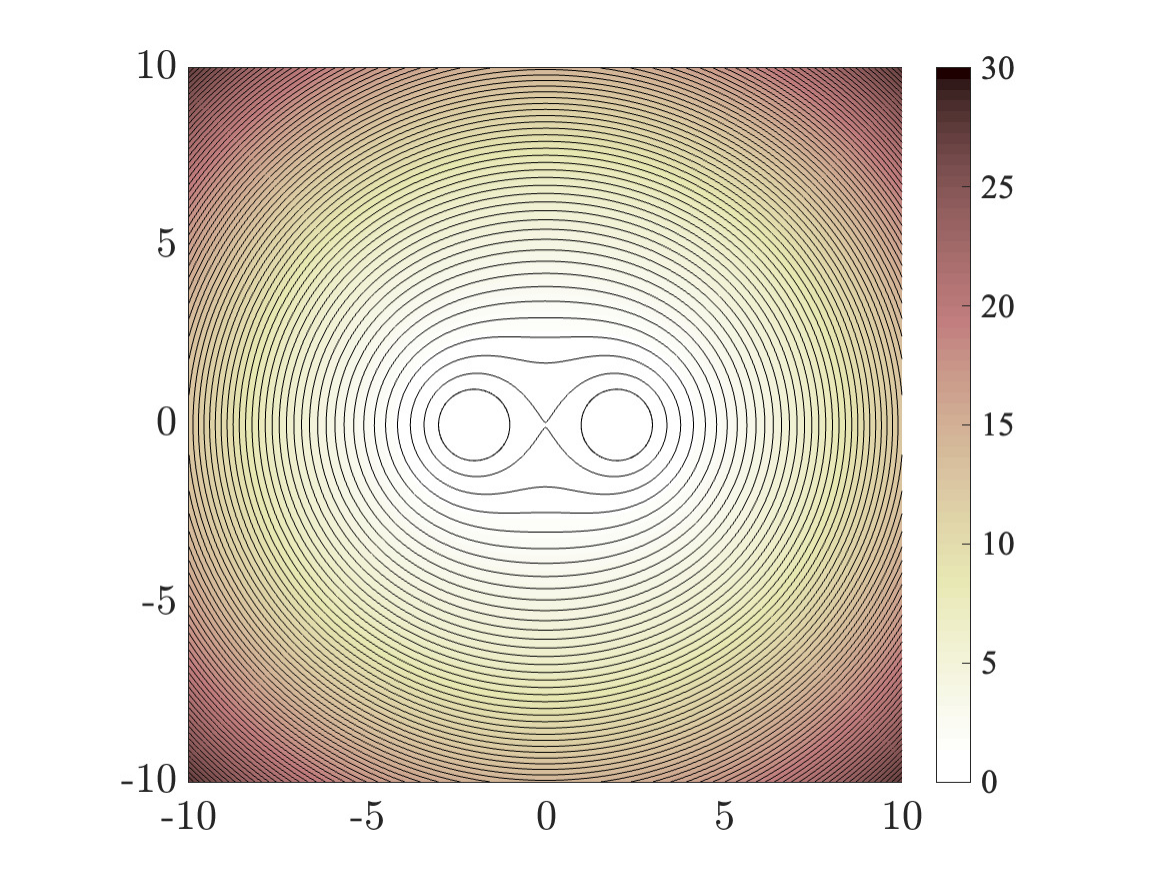}\,
(b)\includegraphics*[height=5cm,trim=14mm 6mm 15mm 0mm,clip]{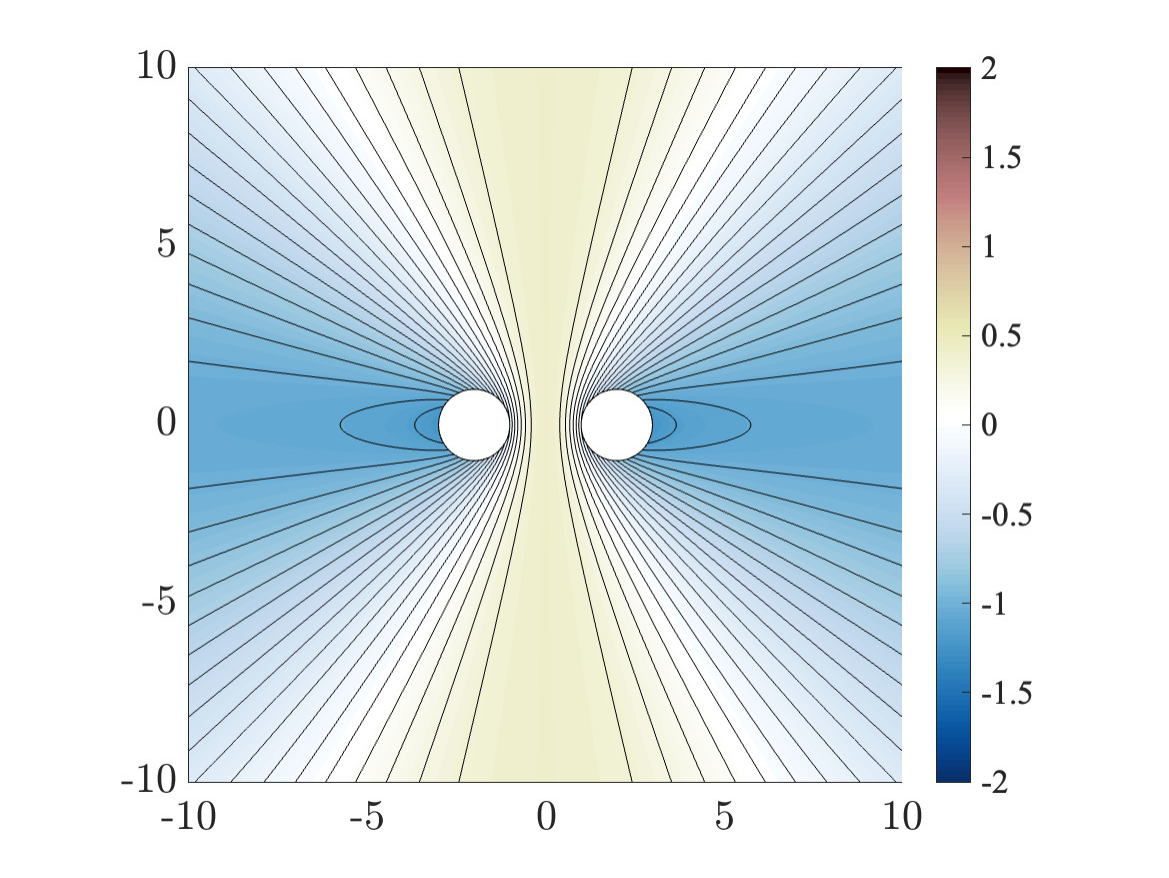}\,
(c)\includegraphics*[height=4.6cm]{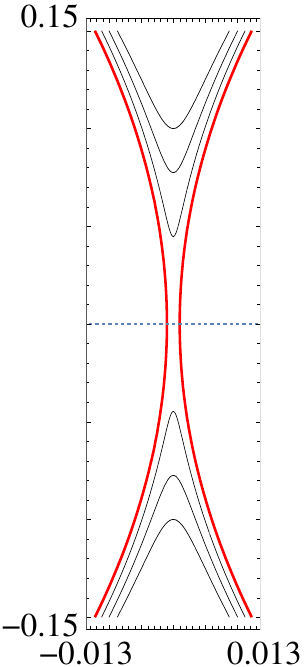}
\end{center}
\caption{  (a) Streamlines  $\psi_{J2}(x,y,\alpha)=$ const. as given by (\ref{Jeffery_streamfunction_2}), for the choice $\epsilon=1\,(\alpha=1.317)$;  the flow asymptotes rapidly to rigid body rotation $\Omega(\alpha)=-0.275$; (b) streamlines in rotating frame, exhibiting  the radial flow in the far field with  $\Lambda(\alpha)=-1.449$; (c) contours $\psi_{L}(x,y)=$ const.~in the narrow gap, for co-rotating cylinders with $\varepsilon=0.001$.}
\label{Streamlinescorot}
\end{figure}

\vskip 5mm \noindent
{\bf 7. Lubrication approach for the two-cylinder problem}
\vskip 2mm \noindent
We consider first the situation when the gap between the cylinders is small ($\varepsilon \ll 1$), and lubrication theory is applicable.  {With the cylinders co-rotating in the clockwise sense ($\Omega_{1}=\Omega_{2}=-1$), the lubrication solution for the vertical component of velocity $v$ is very simple: with  $v=\pm 1$ on the cylinder boundaries  $x=\pm h(y)=\pm(\varepsilon +\thalf y^2)$, this solution is the locally Couette flow $v(x,y) =  x/h(y)$, and the corresponding stream-function satisfying $v(x,y) =-\partial \psi_{L}/\partial x$  is $\psi_{L}(x,y)=\thalf [h(y)^{2}-x^2]/ h(y)$.    The pressure in this Couette-type flow is evidently uniform in the gap}.  
 The streamlines $\psi_{L}=$ const.~are shown in figure \ref{Streamlinescorot}(c) for $\varepsilon=0.001$. The $x$-component of velocity is
\be\label{ulub}
u(x,y) = \frac{\partial\psi_{L}}{\partial y} = y\left[\frac{1}{2}+\frac{2x^2}{(y^2 +2\varepsilon)^2}\right]\,.
\ee
Thus, on the $y$-axis,  $u(0,y) =\thalf y$; it will turn out that this does correctly represent the flow in the lubrication region $y=\tn{O}(\varepsilon^{1/2})$ --- see figure 18(a) below.

The vertical forces on $\mathcal{C}_1$ and $\mathcal{C}_2$ are respectively
\be
\hat{F}_{y}=\mp   \int_{-\infty}^{\infty} \frac{\tn{d}y}{h(y)}= \mp \pi  (2/\varepsilon)^{1/2}\,,
\ee
singular in the limit $\varepsilon\rightarrow 0$.  In this limit these forces are concentrated very near the points $(\mp\varepsilon, 0)$, so the torque on each cylinder about its axis is $\hat{\mathcal{T}}\sim \pi  (2/\varepsilon)^{1/2}$.  However, the torque relative to the origin $(0,0)$ acting on the pair of cylinders is $\hat{G}(\varepsilon)\sim 2 \pi (2\varepsilon)^{1/2} $, vanishing in the limit $\varepsilon\rightarrow 0$. 

\vskip 5mm
\noindent {\bf 8. Jeffery's paradox and Watson's remedy}
\subsubsection{8.1 Jeffery's 1922 solution}
\noindent Using Jeffery's (1922) bipolar coordinates [Appendix A and Fig. 10(b)],
the  solution for co-rotating cylinders ($\Omega_{1}=\Omega_{2}=-1$) may be easily obtained. With 
$\alpha=\cosh^{-1}{(1+\varepsilon)}$, the required streamfunction (an even function of $\xi$ in this case) is found to be
\be\label{Jeffery_streamfunction_2}
\psi_{J2}(x,y) =\frac{\alpha \sinh {\alpha}\cosh{\xi}-\xi \sinh{\xi}\cosh{\alpha}}{h(\xi,\eta)[\alpha +\cosh{\alpha}\sinh{\alpha}]}\,.
\ee
Streamlines for the choice $\varepsilon=1\, (\alpha=1.317)$ are shown in figure  \ref{Streamlinescorot}(a).
For $r\gg 1+\varepsilon$, in polar coordinates and dropping an irrelevant constant, $\psi_{J2}$ has the asymptotic form 
\be\label{psi_J2}
\psi_{J2}(r,\theta)\sim -\thalf\Omega(\alpha) r^2 - \Lambda(\alpha)\cos^{2}\theta+\tn{O}(r^{-2})\,,
\ee
\vskip -5mm
\noindent where
\vskip -5mm
\be
\Omega(\alpha)=-\frac{\alpha}{\alpha +\cosh{\alpha}\sinh{\alpha}}\,,\quad \Lambda(\alpha)=\frac{2\sinh{\alpha}\cosh{\alpha}}{\alpha +\sinh{\alpha}\cosh{\alpha}}\,.
\ee
The first term represents  rigid-body rotation with angular velocity $\Omega(\alpha)$, and, in the frame rotating with this angular velocity, the second term represents the instantaneous far-field radial flow of strength $\Lambda (\alpha)$  apparent in figure  \ref{Streamlinescorot}(b); `instantaneous' because, in this rotating frame, this radial  field  rotates  in tandem with the two cylinders with angular velocity $-\Omega (\alpha)$.  Note that, in the contact limit $\alpha =0$, \,$\Omega (0)=-1/2$ and 
$\Lambda (0)=+1$.
\begin{figure}\begin{center}
\includegraphics*[width=.8\textwidth,  trim=0mm 0mm 0mm 0mm]{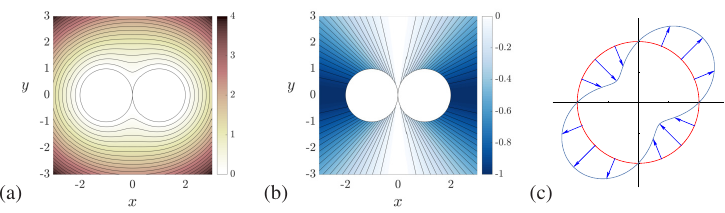} 
 \end{center}
 \caption{(a) Streamlines $\psi(r,\theta)= \tfourth r^2 -\cos^2 \theta=$ const.(in the contact limit $\varepsilon=0$); (b) streamlines $\psi_Q (r,\theta)=-\cos^2 \theta=$ const; (c) velocity profile for the radial quadrupole, defined in \S 8.2. }\label{Jeffery_co_2}\end{figure}

\subsubsection{8.2 Rotating frame solution; radial quadrupole}
\noindent Although the solution (\ref{psi_J2}) tends to rigid-body rotation
$\Omega(\alpha)$ at infinity, it is nevertheless not without interest.  In
a frame of reference rotating with this angular velocity, the cylinders
`orbit' with angular velocity $-\Omega(\alpha)$.  The torque
$\mathcal{T}(\alpha)$ resisting this orbiting is just equal and opposite to the
torque $-\mathcal{T(\alpha)}$ on the pair that is generated by their
co-rotation.  The net torque experienced by the cylinder pair (and equally
the net torque imparted to the fluid) is therefore zero in this situation. In this sense, the situation is quite similar to that of scenario (i) for the model problem in \S6,  when the zero-stress condition is applied on the `remote' cylinder $r=R_{0}$ (which therefore absorbs zero torque).

Figure \ref{Jeffery_co_2}(a) shows the streamlines $\psi_{J2}=$ const.~for
the flow (\ref{psi_J2}) in the contact limit $\varepsilon=0\, (\alpha=0)$, in the frame of the two cylinders. 
Fgure \ref{Jeffery_co_2}(b) shows the streamlines of the same flow in the frame rotating
with angular velocity $\Omega(0)=-\thalf$; these streamlines are again purely radial,
the normal velocity being non-zero on the cylinders because in this
rotating frame the cylinders are orbiting about the origin (as well as
still rotating about their respective axes).  
The streamline pattern of  figure \ref{Jeffery_co_2}(b) corresponds to the streamfunction  
\be\label{radial_quad}
\psi_Q (r,\theta) = -\cos^{2}\theta = -\thalf(1+\cos 2\theta)\,,
\ee
 a very particular solution of the biharmonic equation.  The radial velocity is $u_{r}(r,\theta)= r^{-1}\sin{2\theta}$. In the first quadrant $\{x>0,\,y>0\}$ this may be recognised as the low-Reynolds-number limit of the
 Jeffery-Hamel flow  due to a line source of strength $Q=1$ at the
 intersection of the plane boundaries $x=0,\,y=0$, with no slip on both
 boundaries (\cite{J15}, or see, for example, Batchelor 1967 \cite{B67},
 \S 5.6).  In the second quadrant, it represents  the same flow but with a
 line sink $Q=-1$; and in the third and fourth quadrants, it represents
 again the same flows, source and sink respectively.  The total velocity
 profile going round the circle $r=1$ is as indicated in  figure
 \ref{Jeffery_co_2}(c), outwards in the first and third quadrants,
 inwards in the second and fourth.  It is appropriate to describe this
 flow as a `radial quadrupole'.  

\subsubsection{8.3 Watson's  remedy}
\noindent The flow described by the streamfunction (\ref{psi_J2}) obviously does not satisfy the condition of zero velocity at infinity; this is   `Jeffery's paradox' for the co-rotating situation.  Resolution of this paradox was provided by Watson (1995)\cite{W95}, who compared the situation to the case of a single cylinder rotating in an unbounded fluid  [scenario (ii) of \S6], for which the steady flow is that due to a virtual line vortex at the axis of the cylinder. As Watson remarked ``it seems implausible that the introduction of a second [co-rotating] cylinder would change the character of the motion so drastically" [as to replace this asymptotic vortex flow by a rigid-body rotation].  

\begin{figure}
  \centerline{\includegraphics*[width=0.9\textwidth,  trim=0mm 0mm 0mm 0mm]{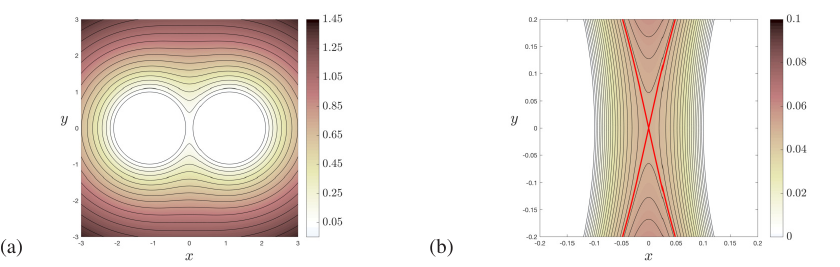}}  
\caption{(a) Contours $\psi_{W}(x,y)=$ const.~as given by (\ref{Watson}) in Appendix B, for the situation when $r_{1}=1,\,\varepsilon=0.1$ (so $\alpha=0.4436$); (b) zoom of the same near the origin, showing in red the expected saddle point where $\psi_{W}(0,0)=0.0491$.}
\label{psiW}
\end{figure}

The essential step in Watson's treatment was to introduce a term proportional to $\log[\{(x-c)^{2}+y^2\}\{(x+c)^{2}+y^2\}]$ in the solution, thereby contributing a vortex ingredient $u_{\theta}=k/r$ to the velocity at infinity and an associated torque $-4\pi\mu k$ acting on the cylinder pair. It was then necessary to satisfy the no-slip conditions on the two cylinders, and to ensure that the rigid-body term is expunged from the solution.  The rather complex details are summarised in Appendix B.     We provide some  diagrams here that help in the interpretation of Watson's results. Denoting his  stream-function by $\psi_{W}(x,y)$, the contours $\psi_{W}(x,y)=$ const.~are shown in figure \ref{psiW}(a) for the case $r_{1}=1,\, \varepsilon=0.1$ (so $\alpha=0.4436$); the zoom near the origin in figure \ref{psiW}(b) shows the expected saddle point at the origin.

\begin{figure}
\begin{center}
\includegraphics*[width=\textwidth,  trim=0mm 0mm 0mm 0mm]{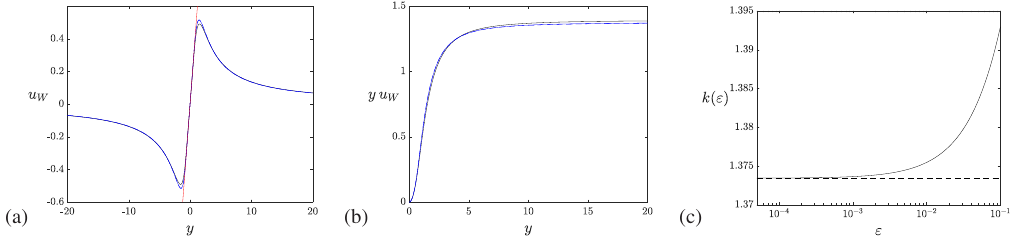}
\end{center}
\caption{(a) Plots of $u_{W}(0,y)$ for  $r_{1}=1$, and $\varepsilon=0.1$ (solid black), $\varepsilon=0.01$ (solid blue), $\varepsilon=0.001$
(dashed blue); the last two curves are indistinguishable on the scale shown. These graphs show the expected behaviour $u_{W}(0,|y|)\sim |y|^{-1}$ for large $|y|$. The red line has slope $1/2$, and correctly represents the flow in the lubrication region $y=\tn{O}(\varepsilon^{1/2})$; 
(b) Corresponding plots of $y\, u_{W}(0,y)$ for positive $y$, showing that in fact  $y\, u_{W}(0,y)\rightarrow k(\varepsilon)$   as $y\rightarrow\infty$.
(c) The function $k(\varepsilon)=\lim_{|y|\rightarrow\infty}|y\,u_{W}(0,y)|$, which asymptotes to  $1.3734$ as $\varepsilon\downarrow 0$; this asymptote is shown by the dashed line, which coincides with the level of
 $\mathcal{T}(0)/4\pi$ as determined by \eqref{asympt_torque}. } 
\label{uW}
\end{figure}

Figure \ref{uW}(a) shows the corresponding velocity $u_{W}(x, y)=\partial\psi_{W}/\partial y$ on the axis $x=0$ for $\varepsilon=0.1$ ($\alpha=0.4436$, solid black), $\varepsilon=0.01$ ($\alpha=0.1413$, solid blue) and  $\varepsilon=0.001$ ($\alpha= 0.04472$, dashed blue); the last two are indistiguishable on the scale shown;  the red line, with slope $1/2$, is as given by the lubrication solution (\ref{ulub}).  Figure \ref{uW}(b) shows that $|u_{W}(y)|\sim k\,|y|^{-1}$ for large $|y|$, as for a point vortex; there is no rigid-body ingredient in this solution. The constant $k$ depends on $\varepsilon$, and may be evaluated numerically as $\lim_{|y|\rightarrow\infty}|y\,u_{W}(y)|$; the function $k(\varepsilon)$ is shown in figure \ref{uW}(c), with the limiting behaviour $k(\varepsilon)\sim 1.3734$ as $\varepsilon\downarrow 0$. 

\begin{figure}
\begin{center}
\includegraphics*[width=.8\textwidth,  trim=0mm 0mm 0mm 0mm]{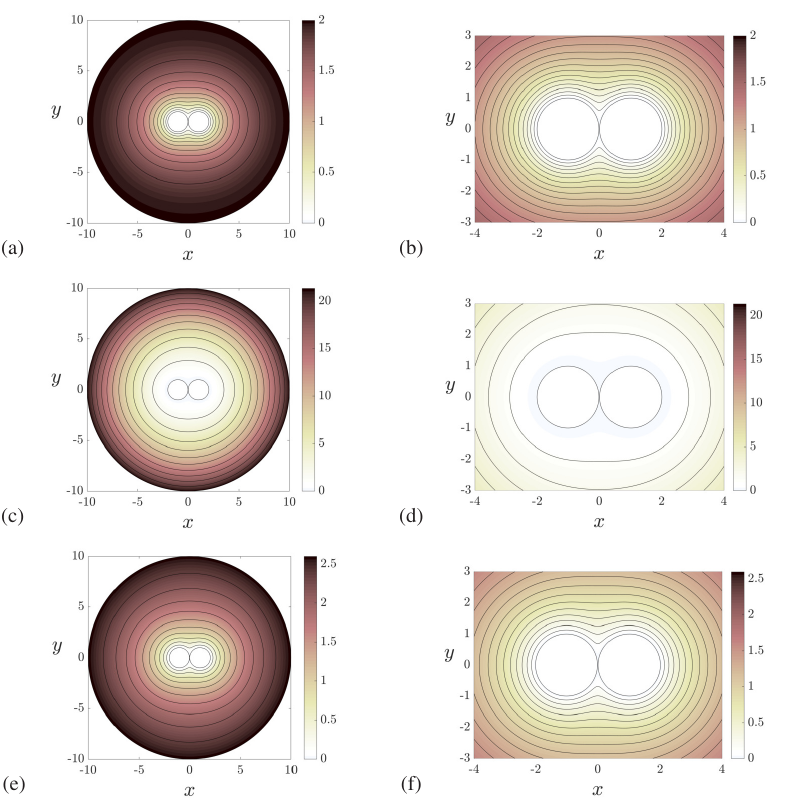} 
\end{center}
\caption{Streamlines $\psi=$ constant from the numerical solution of the
  Stokes problem for co-rotating cylinders with $\varepsilon=0.01$; the outer boundary condition on
  $r=R_{0}$ is either no slip (a,b); stress-free (c,d);  {or (e,f) with $u_{\theta}=k(0.01)/R_{0}$,
from Watson (1995) and figure \ref{uW}(c).}}
 \label{Fig_Corot}
\end{figure}

\begin{figure}
  \centerline{\includegraphics*[width=0.7\textwidth,  trim=0mm 0mm 0mm 0mm]{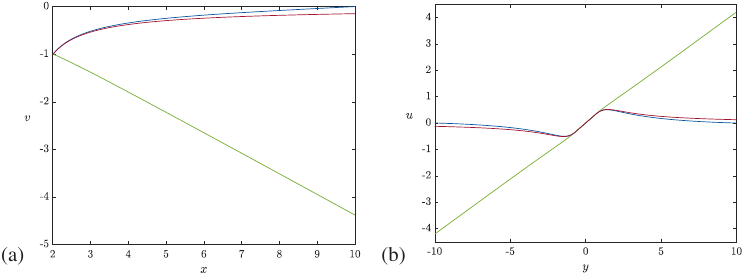}}  
\caption{Axial velocity distributions corresponding to the streamline plots
  of figure \ref{Fig_Corot}; $\varepsilon=0.01$: (a) $v(x,0)$
  ($2+\varepsilon <x<10$); (b) $u(0,y)\, (-10<y<10)$.
The outer boundary condition on
  $r=R_{0}$ is either no slip in blue; stress-free in green; or with
$u_{\theta}=k(0.01)/R_{0}$ in red. The influence of the outer boundary condition is evident: in the no-slip case, the far-field flow is {quite close to that of a point vortex ($\sim r^{-1}$ for $r\gg 1$), except near to $r=R_{0}$ where the no-slip condition is imposed}; in the stress-free case, the far field is rigid body rotation {($\sim r$)}.}
 \label{Fig_Cross}
\end{figure}

Figure \ref{Fig_Corot} shows numerical solutions of the co-rotating problem with $\varepsilon=0.01$ for three different  conditions on the outer boundary $r=R_{0} (=10)$:(a,\,d)  no-slip; (b,\,e) stress-free; and (c,\,f) with the boundary condition $v_{\theta}= -k(\varepsilon)/R_{0}$ [with $k(0.01)$ derived from the Watson solution -- see figure \ref{uW}(c)]. Figure \ref{Fig_Cross} shows corresponding plots of the axial velocity distributions  $v(x,0)$ ($2<x<10$) and $u(0,y)\, (-10<y<10)$.  The Watson solution shows the expected $r^{-1}$ behaviour in both plots [as evident also in Figure 18(b)], and the no-slip solution comes quite close to this.  

The stress-free solution is very different; it shows behaviour close to rigid-body rotation in the far field, as in the solution originally found by Jeffery (1922)\cite{J22}.  Here again we may compare the situation with the model problem introduced in \S 6:  with the stress-free condition [scenario (i)] on the outer cylinder $r=R_{0}$; the steady solution in that case is rigid body rotation no matter how large $R_{0}$ may be.

\subsubsection{8.4 Torque on cylinder pair}
\noindent Having expunged the rigid-body term in the general solution, the asymptotic form of $\psi_W(x,y)$ for $r=(x^2 +y^2)^{1/2}\rightarrow \infty$ is indeed found to be that of a line vortex
\be
\psi_W(x,y)\sim -2K(\alpha) \sinh^{2}\!\alpha \log(1/r)\,,
\ee
where $K(\alpha)$ is the function defined in Appendix B  by (\ref{sum_S}) and (\ref{sum_K}); this
corresponds to a torque  
\be
\hat{\mathcal{T}}(\alpha)= 8\pi K(\alpha) \sinh^{2}\!\alpha
\ee
exerted by the cylinder pair on the fluid.  For small $\varepsilon\sim \thalf\alpha^{2}$, using the asymptotic form (\ref{K_asymp}) (see Appendix B), this implies that
\be
\hat{\mathcal{T}}(\alpha)\sim 8\pi \alpha^{2} (0.6867 \alpha^{-2}) = 17.2587\quad\tn{as}\,\,\alpha\rightarrow 0\,;  \quad \tn{so} \quad \hat{\mathcal{T}}(0)/4\pi=1.3734,
\label{asympt_torque}
\ee
(cf. $\hat{\mathcal{T}}_{s}/4\pi = 1$ for a single rotating cylinder). The dashed line in figure  \ref{uW}(c) shows that 
$\hat{\mathcal{T}}(0)/4\pi=1.3734$ exactly as given by (\ref{asympt_torque}), and the agreement is evidently excellent. Just as for a single cylinder, this torque arises from the stress distribution  around the entire  boundary of the two cylinders, and the result cannot therefore be obtained from the lubrication theory of \S 7.

\begin{figure}
\begin{center}
\includegraphics*[width=0.35\textwidth,  trim=0mm 0mm 0mm 0mm,clip]{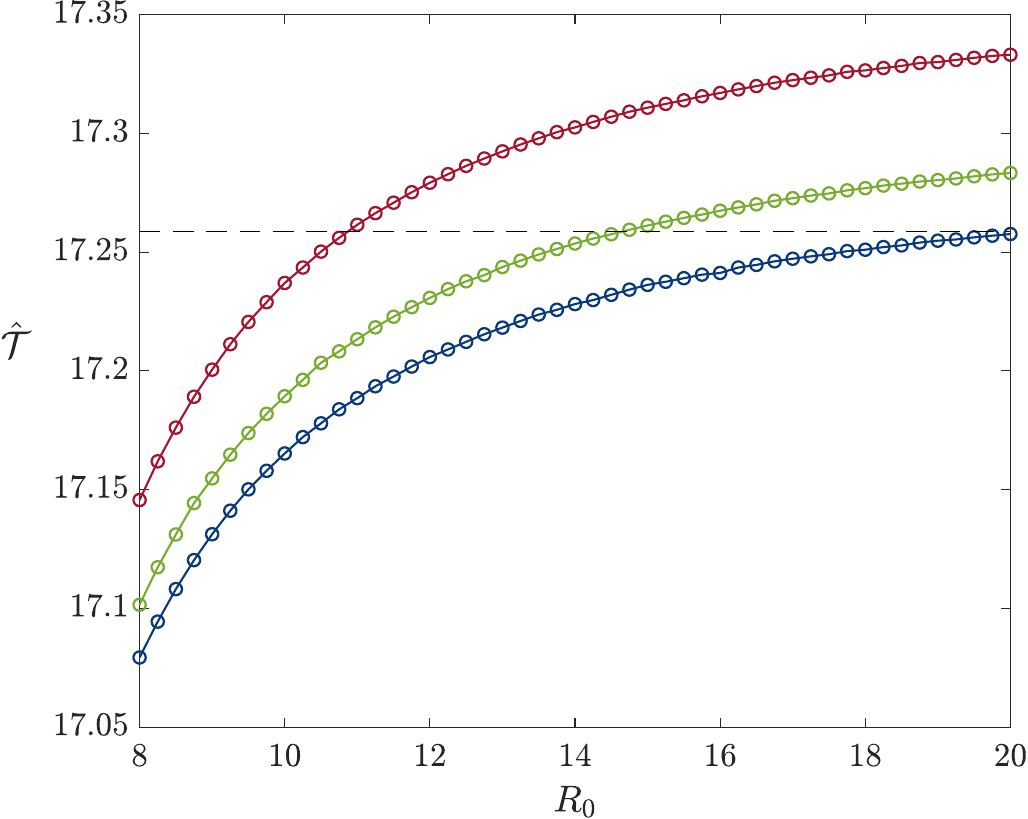} 
\end{center}
\caption{Torque on the  bounding cylinder $\mathcal{C}_{0}$ (with no-slip boundary condition) as a function of $R_{0}$ for
  $\varepsilon = 0.04$ (red),
  $0.02$ (green),
  $0.01$ (blue).
  The double limit 
  $R_{0}\rightarrow \infty$ and  $\varepsilon\downarrow 0$  {(taken numerically in either order)} is needed to approach the
  asymptotic value 17.2587, marked by the dashed line, as derived in (\ref{asympt_torque}).}
 \label{Fig_Seventeen}
\end{figure}

In the finite-domain numerical solution, the torque exerted on the fluid jointly by the two inner cylinders must equal the torque absorbed by the outer cylinder $\mathcal{C}_{0}$.  Figure \ref{Fig_Seventeen} shows the dependence of this torque on $R_{0}$ (with the no-slip boundary condition) in the range $8 \le R_{0}\le 20$, for $\varepsilon= 0.04,0.02$ and $0.01$;  here, it is the double limiting process  $R_{0}\rightarrow \infty, \varepsilon\downarrow 0$ that shows convergence towards the limit 17.2587, again marked by the dashed line.

\section{Discussion and Conclusions}\label{conclusion}

We have here  reinvestigated the classical problem of the Stokes flow generated by two parallel rotating circular cylinders of unit radius.  We have assumed that the fluid domain $\mathfrak{D}$ is bounded by a cylinder $r=R_{0}\gg 1$ on which either no-slip or stress-free boundary conditions can be imposed, and have provided an accurate finite-element numerical treatment of this problem in two cases: counter-rotating cylinders in Part I (\S\S 1--5), and co-rotating cylinders in Part II (\S\S6--8).  Note that, insofar as the Stokes problem is linear, the case of arbitrary cylinder  rotation rates $\Omega_{1}$ and  $\Omega_{2}$  is just a linear combination of these two prototypical cases.
\vskip 1mm
\noindent \emph {Counter-rotating cylinders}
\vskip 1mm\noindent 
 A  model problem exhibiting the same symmetries as the two-cylinder problem was first analysed in 
\S 2. This allows asymptotic treatment in an inner regime $1<r\ll R_{0}$, and in an outer regime 
$\{r= k R_{0},\, k<1,\, k=\tn{O}(1)\}$ where the influence of inertia forces may be estimated. This shows that, no matter how small the Reynolds number may be, inertia always becomes significant in the outer regime as $[\log{R_{0}}]$ increases without limit. 

 When the gap $2\varepsilon$ between the inner cylinders  is small, lubrication theory provides a description of the flow in the gap region that agrees well with the numerical solution. When $\varepsilon\downarrow 0$,  two saddle points at $(0,\pm (6\varepsilon)^{1/2})$ move towards the origin.  When the cylinders make contact (i.e.~when $\varepsilon=0$), the domain topology changes from triply- to doubly-connected, and the two saddle points convert to two boundary-saddle points.  The numerical evidence indicates that when $\varepsilon>0$ the force exerted on the cylinders tends to zero as $R_{0}\rightarrow \infty$; this conclusion is consistent with  the  model calculation of \S 2,  the lubrication analysis of \S 3.2, and the analytical solution (for $\varepsilon=0$) obtained in \S 4.  In this case, there is a `contact force' due to the discontinuity of pressure across the point of contact, and we have shown that the total force is continuous down to $\varepsilon=0$ if and only if this contact force is taken into account. 

 Based on our results, we may envisage two distinct scenarios for this counter-rotating situation when the fluid is unbounded.  If the cylinders are free of any external restraining force, they will  move relative to the fluid, ultimately with uniform dimensionless velocity $V \left[=\thalf (1+\varepsilon)^{-1}\right]$ in the $y$-direction (just like two parallel line vortices of circulations $\pm \Gamma$ separated by distance $2d$ in an inviscid fluid, which propagate with velocity $V=\Gamma/4\pi d$). In a frame of reference moving with the cylinders, there is thus a uniform streaming velocity $(0,-V)$ at infinity.  The ultimate steady state is therefore well described by Jeffery's original solution, the perturbation from the uniform stream being the `torquelet' as given by (\ref{psi_polars}).  The velocity associated with the torquelet is O$(r^{-2})$ as $r\rightarrow \infty$, so that ${\bf u}^2 =$O$(r^{-4})$. The kinetic energy of the flow in the frame of reference in which the fluid is at rest at infinity is therefore finite, and this flow  could therefore be generated from a state of rest in a finite time.  
 
 If however, the cylinders are constrained in such a way that their axes are fixed as they rotate, the situation is quite different. If  in this situation the rotation is started impulsively at $t=0$, then a low Reynolds number flow will develop, as analysed by Ueda et al.~(2003)\cite{U2003}.  This flow incorporates both stokeslet and  uniform-streaming ingredients, but extends only over a range 
 $r\lesssim \tn{Re}^{-1}\log{[\tn{Re}^{-1}]}$,  limited by inertia effects. The cylinders must then experience a net force  $\hat{F}_{y}$ in the limit $\tn{Re}\rightarrow 0$ that we estimate to be $\hat{F}_{y}\sim 1/\log{[\tn{Re}^{-1}\log{[\tn{Re}^{-1}]}}$ . The problem of matching this Stokes flow to an outer solution of the full Navier-Stokes equations remains open.

\vskip 1mm
\noindent \emph {Co-rotating cylinders}
\vskip 1mm\noindent 
Again in this case, an elementary  model problem is presented in \S 6, for which the boundary condition on the outer cylinder $r-R_{0}$ may be stress-free or no-slip. For the two-cylinder co-rotating problem, these alternatives correspond to two quite distinct scenarios which emerge from our analysis,  We have provided a finite-element numerical solution for both scenarios when  $R_{0}\gg 1$.  Lubrication theory describes well the highly sheared flow in the narrow gap when $\varepsilon\ll 1$, and  good agreement is achieved with the full numerical solution.

In the first scenario, a zero-stress condition is applied on $r=R_{0}$, and the flow tends in the far field to rigid-body rotation with angular velocity $\Omega(\alpha)$, where $\alpha= \cosh^{-1}(1+\varepsilon)$; there is in this situation zero net torque on the cylinder pair.  In an unbounded fluid, this is essentially the solution of Jeffery (1922). In a frame of reference rotating with the fluid at infinity the cylinders orbit with angular velocity $-\Omega(\alpha)$, and this is the behaviour if their axes are unconstrained.  In this rotating frame, the fluid velocity is identified as a `radial quadrupole'.

In the second scenario, the cylinder axes are supposed be held fixed by external constraints.  Their co-rotation about these fixed axes  then generates a localised torque on the fluid and so a vortex-like flow $u_{\theta} \sim k(\varepsilon)r^{-1}$ in the far field.  [This is approximately  realised numerically by imposing a no-slip condition on $r=R_{0}$.]  There is also an internally driven rigid-body rotation, but, following Watson (1995), this is nullified by suitable choice of the strength of  the virtual vortex [thus in effect determining the function $k(\varepsilon)$], so that the resulting velocity tends to zero for large $r$ thus resolving Jeffery's paradox in this case.  We have extended and refined  Watson's analysis to reveal the behaviour as $\varepsilon\downarrow 0$; in this limit we have shown that the torque $\hat{\mathcal{T}}$ on the cylinder pair asymptotes to the value 17.2587.

\vskip 2mm

\appendix

\vskip 2mm

\appendix

\section{Appendix A: Asymptotic behaviour of Jeffery's (1922) solution}\label{App_Jeffery}
\noindent Jeffery(1922) used bipolar coordinates defined by a conformal mapping 
$\xi +i\eta =  \log{\frac{x+iy+c}{x+iy-c}}$,  so that
\be\label{bipolar_2}
 \xi(x,y)=\thalf \log{\frac{(x+c)^2 +y^2}{(x-c)^2 +y^2}}\,,\quad  \eta(x,y)=
\tan^{-1}\!\left[\frac{x\!+\!c}{y}\right]-\tan^{-1}\!\left[\frac{x\!-\!c}{y}\right]\,.
\ee
The scale factor for this mapping is 
$h(\xi,\eta)=\left|d\zeta/dz\right|=(\cosh\xi - \cos\eta)/c$.
The contours $ \xi(x,y)=$ const., $ \eta(x,y)=$ const.  are shown in figure \ref{contours}(b). The contour $\xi=\alpha$ is a circle with centre at $(c_{1}, 0)$ where $c_{1}=c\, \coth{\alpha}$, and radius  $r_{1}=c\, \tn{cosech} |\alpha|$. If we fix $r_{1}=1$, then $c=\sinh{|\alpha|}$, and the separation of the two cylinders is (as before)  $2\varepsilon$, where now 
$\varepsilon= \cosh{\alpha}-1$, i.e.~$ \alpha =\cosh^{-1}(1+\varepsilon)$.
The boundaries $\mathcal{C}_{1,2}$ are taken to be $\xi=\mp \alpha$ with $\alpha>0$, and the fluid domain is 
$\mathfrak{D}: \{0\le |\xi|<\alpha,\,-\pi<\eta<\pi\}$.

With this notation, Jeffery (1922, p.173) found the streamfunction for the counter-rotating situation in the form 
\be\label{Jeffery_streamfunction}
\psi_{J}(x,y) = [h(\xi,\eta)]^{-1}\left[b_{0} \,\xi \,(\cosh{\xi}-\cos{\eta})+c_{0}\sinh{\xi}+c_{1}\sinh{2\xi}\cos{\eta}\right]\,,
\ee
where the constants $b_{0},  c_{0}$ and $c_{1}$ take values determined by the boundary conditions 
\be
b_{0}=-\frac{\cosh{2\alpha}}{2\cosh{\alpha}\sinh^{2}{\alpha}}\,,\quad c_{0}=\frac{\cosh{\alpha}}{2\sinh^{2}{\alpha}}\,,\quad
c_{1}= -\frac{1}{4\cosh{\alpha}\sinh^{2}{\alpha}}\,.
\ee
The streamfunction $\psi_{J}$ then takes the asymptotic form (for $r= (x^2 + y^2)^{1/2}\gg 1$)
 \be\label{psiJ1}
\psi_{J}(x, y)\sim \frac{x}{2 \cosh {\alpha}} -\frac{2x\left[x^{2}+(x^{2}+y^{2})\sinh ^{2}{\alpha}\right]}{\cosh{\alpha}\,(x^{2}+y^{2})^{2}} +\tn{O}(r^{-3})\,,
\ee
or equivalently in polar coordinates
 \be\label{psiJ2}
\psi_{J}(r,\theta)\sim \frac{r \cos\theta}{2 \cosh\alpha} -\frac{(4\cosh^{2}\alpha-1)\cos\theta +\cos{3\theta}}{2r\cosh\alpha} +\tn{O}(r^{-3})\,,
\ee
giving (\ref{psi_polars}) in the limit $\alpha=0$. 
The leading term in (\ref{psiJ1}) represents a uniform stream 
$(0, -\thalf \tn{sech} \,\alpha)\!=\![0, -\thalf(1\!+\!\varepsilon)^{-1}]$ as found by Jeffery,  and the second term, of order $r^{-1}$, which has not been previously identified, represents a torquelet, whose strength, noting the structure of (\ref{psiJ2}), may be defined as
\be
\mu_{\mathcal T} =\pi (4\cosh^{2}\alpha-1)\tn{sech}\, \alpha = d(\alpha)\, \hat{\mathcal{T}}_{2}(\alpha)/2\quad\tn{say}.
\ee
Here, $d(\alpha)$ is the separation of virtual point torques of magnitude $\pm \hat{\mathcal{T}}_{2}(\alpha)$ needed to give precisely this torquelet strength.  The flattening of the streamlines evident in figure \ref{torqueletf}(a) decreases as $\alpha$ increases,
 i.e.~as the separation of the cylinders is increased, with corresponding decrease of their mutual interaction.

\begin{figure}
\begin{center}
(a)\includegraphics*[height=3.2cm, trim=0mm 0mm 0mm 0mm, clip]{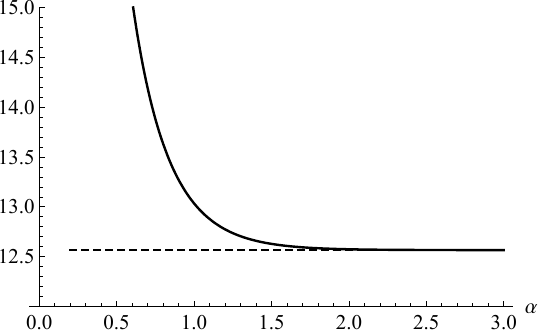}\quad
(b)\includegraphics*[height=4.4cm, trim=15mm 3mm 18mm 0mm, clip]{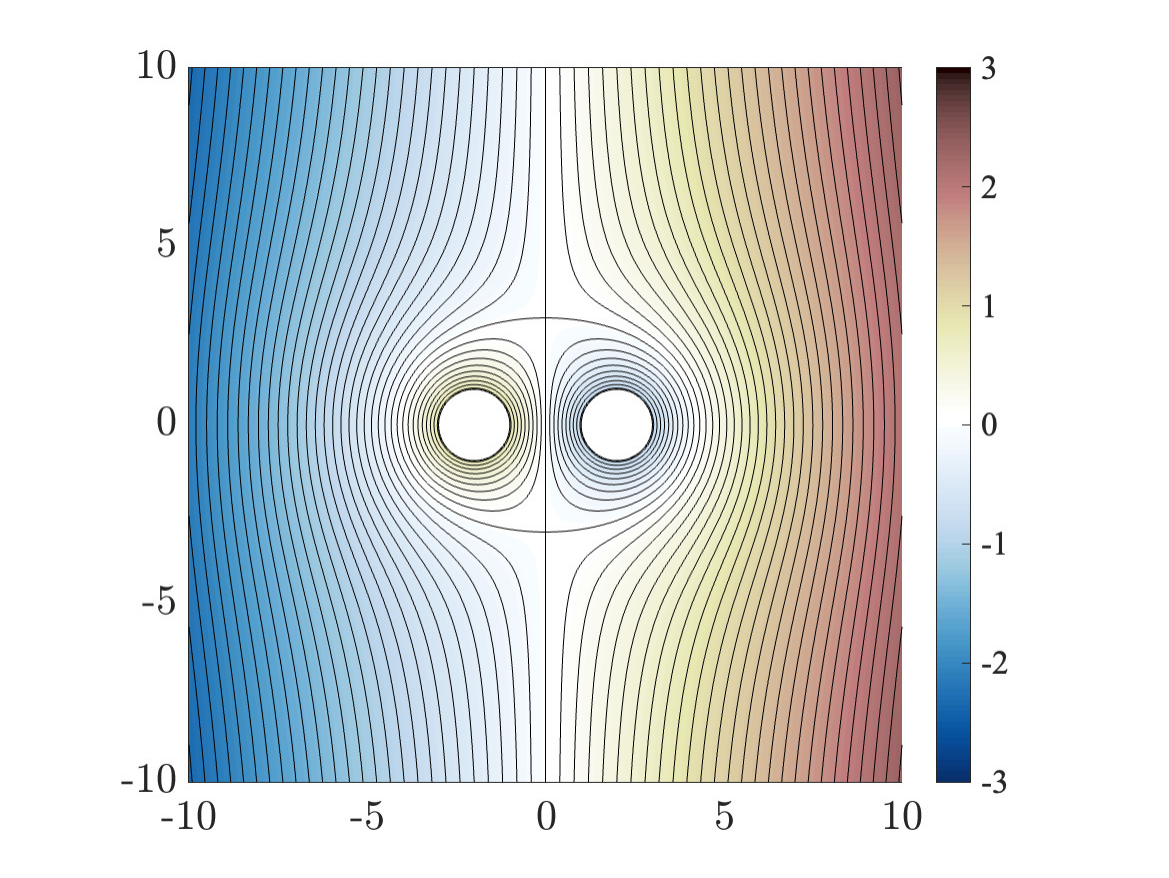}\quad
(c)\includegraphics*[height=3.2cm, trim=0mm 0mm 0mm 0mm, clip]{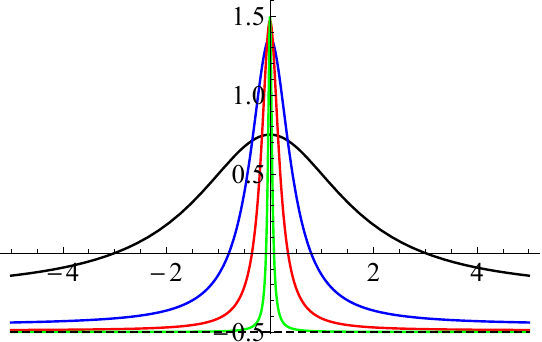} 
\end{center}
\caption{(a) The function $\hat{\mathcal{T}}_{2}(\alpha)$ (cf.~figure\ref{Torque_J}(a)) showing the large-$\alpha$ asymptote (dashed) at the level $4\pi\approx12.566$; (b) streamlines  $\psi_{J}(x,y,\alpha)=$ const. as given by (\ref{Jeffery_streamfunction}), for the choice $\varepsilon=1\, (\alpha=1.317)$; for this choice, the saddle points are at $y=\pm 3$ and the flow asymptotes to the uniform stream $(-\tfourth,0)$ as $r\rightarrow\infty$; (c) the velocity $v_{J}(0,y)$ for $\varepsilon=$ 1 (black), 0.1 (blue), 0.2 (red) and 0.001 (green); in the limit $\varepsilon =0$, the asymptotic uniform stream is $(-\thalf,0)$, as shown by the dashed line.}
\label{Jeffery_co}
\end{figure}
Figure \ref{Jeffery_co}(a)
again shows the function $\hat{\mathcal{T}}_{2}(\alpha)$ computed from Jeffery's solution (\ref{Jeffery_streamfunction}), but here in the range 
$\alpha\gtrsim 0.5,\,\,(\varepsilon\gtrsim0.13)$; this shows rapid approach to the asymptotic level $4\pi$, this limit being quite accurately reached for $\alpha\gtrsim 2$, i.e.~for $\varepsilon=\cosh\alpha -1 \gtrsim 2.75$. The separation $d(\alpha)$ then asymptotes to $2\cosh\alpha=2(1+\varepsilon)$, the distance between the centres of $\mathcal{C}_{1}$ and $\mathcal{C}_{2}$,  as might be expected. 

At the other extreme  $\alpha\rightarrow 0$, as we have seen from figure \ref{Torque_J}(a), $\hat{\mathcal{T}}_{2}(\alpha)\sim 2\pi/\alpha$, so that in this limit the separation $d(\alpha)$ is given by
$d(\alpha)  \sim 6\pi/ \hat{\mathcal{T}}_{2}(\alpha) \sim 3\, \alpha \sim 3\,(2\varepsilon)^{1/2}$,
as stated in \S 3.4.  For  $\alpha \sim (2\varepsilon)^{1/2}\ll 1$ \,(and $r\gg \varepsilon^{1/2}$, i.e.~outside the lubrication zone), the solution (\ref{Jeffery_streamfunction}) may be expanded in powers of $\varepsilon$; in polar coordinates, at leading order this gives
\be\label{jeff_1}
\psi_{J}(r,\theta)= \thalf r\cos\theta -\thalf r^{-1}(3 \cos\theta +\cos 3\theta)+\tn{O}(\varepsilon)\,,
\ee
recovering the result (\ref{psi_polars}) when $\varepsilon=0$.

At the intermediate value $\varepsilon =1\, (\tn{so}\,\, \alpha=1.317)$,  the streamlines $\psi_{J}=$ const. are as shown in figure \ref{Jeffery_co}(b).  For this choice of  $\varepsilon$, the saddle points are at $y=\pm 3$ and the flow settles to the uniform stream $(0,-\tfourth)$ as $r\rightarrow\infty$. Moreover, $ \hat{\mathcal{T}}_{2}(1.317)=12.70$ and $d$ evaluates to $3.71$, somewhat less than $4 \,[=2(1+\varepsilon)]$, the distance between the centres of $\mathcal{C}_{1}$ and $\mathcal{C}_{2}$. This is to be expected, because the tangential stress $\hat{\tau}(\phi)$ on $\mathcal{C}_2$ becomes more pronounced near the gap location $\phi=\pm \pi$ as $\varepsilon$ decreases (and similarly of course for $\mathcal{C}_1$).

Figure \ref{Jeffery_co}(c) shows the velocity
 $v_{J}(0,y)$ for the four choices  $\varepsilon= 1,\,0.2,\,0.1$ and $0.001$. For large $|y|$,
\be
 v_{J}(0,y) \sim -\frac{1}{2\cosh \alpha} +\frac{2 \sinh ^{2}\alpha}{\cosh\alpha} \,y^{-2} +\tn{O}(y^{-4})= 
 -\frac{1}{2(1+\varepsilon)}+\frac{2\varepsilon(2+\varepsilon)}{(1+\varepsilon)} \,y^{-2} +\tn{O}(y^{-4}),
\ee  
the leading term giving the streaming velocity  $\left[0,-\thalf(1+\varepsilon)^{-1}\right]$ as $|y|\rightarrow\infty$.

 \section{Appendix B: Asymptotic analysis of Watson's (1995) solution}\label{App_Watson}
\begin{figure}
\begin{center}
\raisebox{0.1cm}{(a)}\includegraphics*[height=4cm]{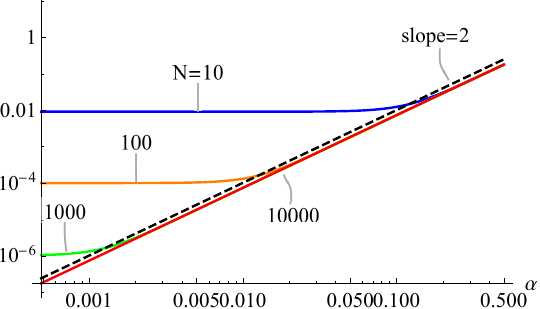} \qquad
\raisebox{0.1cm}{(b)}\includegraphics*[height=4cm]{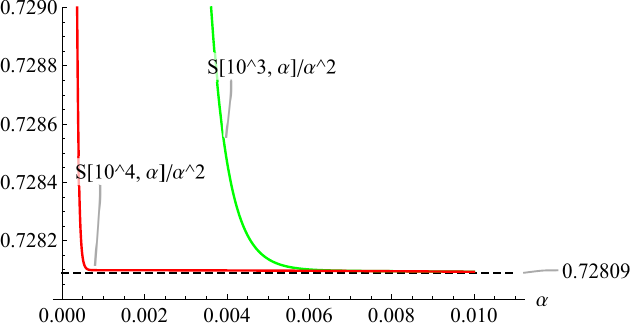} 
\end{center}
\caption{(a) Log-log plots of the functions $S[N,\alpha]$ for $N=10,10^2,10^3,10^4$, indicating that $S[N,\alpha]\sim k\,\alpha^2$ over a range of $\alpha$ that extends towards $\alpha=0$ with increasing $N$ ; (b) plots of $\alpha^{-2}S[N,\alpha]$ for $N=10^3,10^4$, indicating that  $k\approx 0.7281$. }
\label{S_asymptotic}
\end{figure}

\noindent  Watson's (1995) solution of the co-rotating problem involved the function
\be\label{sum_S}
S(\alpha)=\thalf +\frac{\alpha\sinh^{2}\alpha\tanh\alpha}{\alpha+\sinh\alpha\cosh\alpha} 
-4\sum_{n=2}^{\infty} \frac{ \sinh{\alpha} (\sinh\alpha+n  \cosh\alpha)+n\, e^{-n\alpha} \sinh{n\alpha}}{(n^{2}-1)(n\sinh{2\alpha}+\sinh{2n\alpha})}\,,
\ee
where $\alpha$ is related  to the gap parameter $\varepsilon$ (as in Appendix A) by $\alpha=\cosh^{-1}{(1+\varepsilon)}$.
The convergence of the sum here is an issue of immediate concern.  For any fixed $\alpha>0$, the coefficient in the $n$th term of the sum is proportional to $n^{-1}\,e^{-2n\alpha}$ for large $n$, so that convergence is assured.  However, as $\varepsilon\rightarrow 0$, i.e.~ $\alpha\rightarrow 0$, more and more terms of the series must be retained to get any prescribed level of  accuracy.  Defining $S[N,\alpha]$ as the function (\ref{sum_S}) when the summation is truncated at $n=N$, figure \ref{S_asymptotic}(a) shows  log-log plots of $S[N,\alpha]$, for $N=10,10^2,10^3,10^4$.  The dashed line has slope 2, indicating that, as $N\rightarrow \infty$, $S[N,\alpha]\sim k\,\alpha^2$ for some constant $k$.  Figure \ref{S_asymptotic}(b) shows the functions $S[10^{3},\alpha]/\alpha^{2}$ and $S[10^{4},\alpha]/\alpha^{2}$, indicating that  $k=0.7281$ to good approximation. As expected, this behaviour breaks down when $\alpha$ is too small ($\lesssim 0.001$ when $N=10^4$), but  it nevertheless indicates that the limiting function $S(\alpha)$ has an asymptotic behaviour 
\be\label{S_asym}
S(\alpha) \sim 0.7281\, \alpha^2 \quad \tn{as}\,\,\alpha\rightarrow 0\,.
\ee
A second function $K(\alpha)$ is then defined by
\be\label{sum_K}
K(\alpha) =\frac{\alpha}{S(\alpha)(\alpha +\sinh \alpha\cosh\alpha)}\,,
\ee
with asymptotic behaviour that follows from (\ref{S_asym}),
\be\label{K_asymp}
K(\alpha) \sim 0.6867 \alpha^{-2} \quad\tn{as}\,\,\alpha\rightarrow 0\,;
\ee
(and actually $\alpha^2 K(\alpha) < 0.6868$ for all $\alpha > 0$). 
\begin{figure}
\begin{center}
(a)\ \includegraphics*[height=4cm]{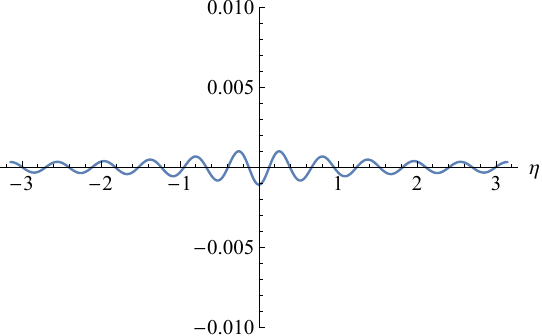} \qquad
(b)\ \includegraphics*[height=4cm]{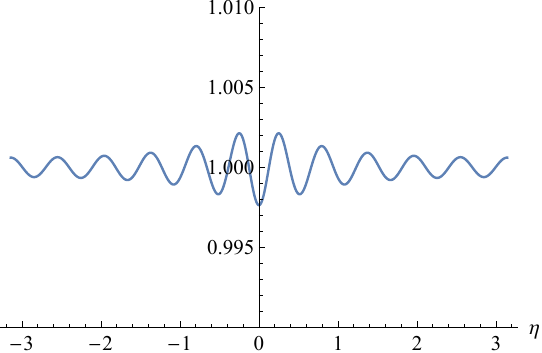} \\[3mm]
(c)\ \includegraphics*[height=4cm]{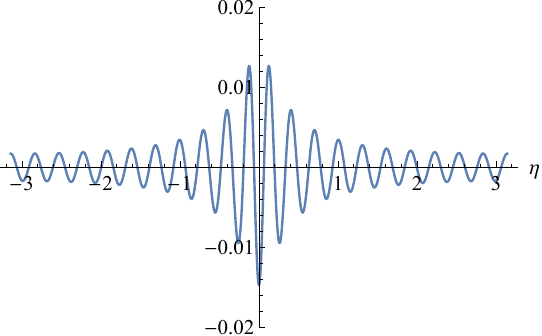} \qquad
(d)\ \includegraphics*[height=4cm]{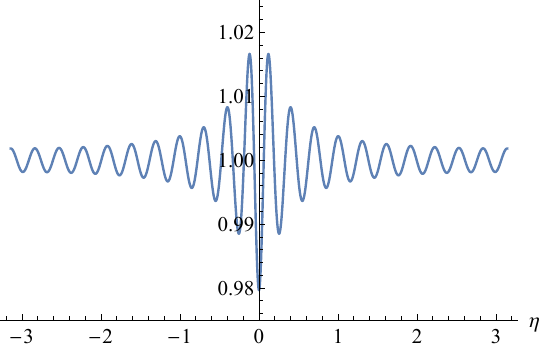}\\
\end{center}
\caption{Verification of the impermeability and no-slip conditions (\ref{imper}) on $\xi=\alpha$; (a) and (b): 
$ \phi(\alpha,\eta)/ \phi(0,\pi)$ and  
$\sinh\alpha\, \partial\phi/\partial\xi|_{\xi=\alpha}$; \,  $\alpha=0.4583, N=10,  \phi(0,\pi)=-0.4681$; (c) and (d): the same for $\alpha=0.1413, N=20,\phi(0,\pi)=-0.4972$.}
\label{noslip}
\end{figure}

In view of the symmetry, we may restrict attention to the half-space $x\ge 0$, and to the fluid region $0\le \xi < \alpha$. With $\xi(x,y)$, $\eta(x,y)$ and $h(\xi,\eta)$ as defined in Appendix A of Part 1, Watson's streamfunction is  given by
\be\label{Watson}
\psi_{W}(x,y)= -\frac{\sinh{\alpha} \,\,\,\phi[\xi(x,y),\eta(x,y)]}{h[\xi(x,y),\eta(x,y)]}\,,
\ee
where $\phi(\xi,\eta)$ has the Fourier series representation
\bea\label{sum_xi}
\phi(\xi,\eta)= K(\alpha)(\cosh\xi\!\!&-&\!\!\cos\eta)\log(2\cosh\xi-2\cos\eta)+a_{0}(\alpha)\cosh\xi+b_{0}(\alpha)\xi\sinh\xi\nonumber\\
&+&\sum_{n=1}^{\infty} [a_{n}(\alpha)\cosh(n\!+\!1)\xi +b_{n}(\alpha)\cosh(n\!-\!1)\xi]\cos n\eta\,.
\eea
Here the coefficients  are given by
\be\label{ab0}
a_{0}(\alpha)= -\frac{\alpha+K(\alpha)(\alpha +\alpha^2 +e^{-\alpha}\sinh\alpha)}{\alpha+\sinh\alpha\cosh\alpha},\quad 
b_{0}(\alpha)=\frac{\coth\alpha-K(\alpha)\sinh^{2}\alpha}{\alpha+\sinh\alpha\cosh\alpha},
\ee
\be\label{ab1}
a_{1}(\alpha)= \thalf K(\alpha)e^{-\alpha} \tn{sech}\,\alpha,\quad b_{1}(\alpha)=K(\alpha)(1+\alpha-\thalf \tanh\alpha),
\ee
and, for $n\ge 2$,
\be\label{ab_n}
a_{n}(\alpha)\!=\!\frac{2K\!(\alpha)(n e^{-\alpha}\sinh\alpha \!+\!e^{-n\alpha}\sinh n\alpha)}{n(n+1)(\sinh 2n\alpha +n\sinh 2\alpha)}\,,\,\,
b_{n}(\alpha)\!=\!-\frac{2K(\alpha)(n e^{\alpha}\sinh\alpha \!+\!e^{-n\alpha}\sinh n\alpha)}{n(n-1)(\sinh 2n\alpha +n\sinh 2\alpha)}\,.
\ee
Again, for any fixed $\alpha>0$, these coefficients have the asymptotic behaviour 
\be
a_{n}(\alpha) \sim \frac{4 K(\alpha)e^{-\alpha} \sinh\alpha}{n\, e^{2n\alpha}}\,,\quad b_{n}(\alpha) \sim -\frac{4 K(\alpha)e^{\alpha} \sinh\alpha}{n \,e^{2n\alpha}}\qquad \tn{as}\,\, n\rightarrow\infty\,,
\ee
so that the coefficient of $\cos {n\eta}$ in (\ref{sum_xi}) has the behaviour
\be
[ a_{n}(\alpha)\cosh(n+1)\xi +b_{n}(\alpha)\cosh(n-1)\xi] \sim
\frac{4K(\alpha)}{ n\,e^{ (n-1)(\alpha-\xi)}\,e^{n\alpha}}\,\quad \tn{as}\,\, n\rightarrow\infty\,.
\ee
The series in (\ref{sum_xi}) therefore converges exponentially rapidly for $0\le \xi \le \alpha$, and may again be evaluated to good approximation if terminated at $n=N$ provided  $N$ is appropriately chosen in relation to $\alpha$.  Since $\alpha\sim(2\varepsilon)^{1/2}$ for $\varepsilon\ll 1$, we therefore need $N\gg(2\varepsilon)^{-1/2}$. When $\varepsilon=0.1$, this requirement is simply $N\gg 1$; we actually ran to $N = 1000$ in constructing figures \ref{psiW} and \ref{uW}, thus providing extreme accuracy. If $N=10$ is used instead, the resulting figures are just as good at this value of $\varepsilon$, but if $\varepsilon$  is reduced, $N$ must be correspondingly increased to maintain accuracy.

The definitions (\ref{ab0}) -- (\ref{ab_n})  ensure that the impermeability and no-slip conditions are satisfied on the  co-rotating cylinders $\xi=\pm\alpha$. These conditions are
\be\label{imper}
\phi(\alpha,\eta)= 0\quad \tn{and}\quad \sinh\alpha\,\left.\frac{\partial\phi}{\partial\xi}\right|_{\xi=\alpha}= 1\quad\tn{for}\,\,-\pi<\eta\le\pi\,.
\ee
Normalising $\phi(\alpha,\eta)$ by its value at $\xi=0,\,\eta=\pi$ (corresponding to the point $x=0,\,y=0$), figure (\ref{noslip}) shows  $\phi(\alpha,\eta)/ \phi(0,\pi)$ and $\sinh\alpha\,\, \partial\phi/\partial\xi|_{\xi=\alpha}$, evaluated for  two cases: 
$\varepsilon=0.1\, (\alpha=0.4583)$ truncating the series at  $N=10$; and  $\varepsilon=0.01\, (\alpha=0.1413)$ truncating the series at $N=20$. The conditions are satisfied to well within $0.2\%$  and $2\%$ accuracy respectively at these levels of truncation.  As $\varepsilon$ is further decreased, $N$ must be correspondingly increased to retain accuracy.

 The coefficient of the `rigid-body' term in the solution as $r\rightarrow \infty$ (i.e.~as $\xi^2+\eta^2\rightarrow 0$) is proportional to
\be\label{sum_R}
R(\alpha)\equiv a_{0}(\alpha) +\sum_{n=1}^{\infty}[a_{n}(\alpha) +b_{n}(\alpha) ]\,,
\ee
and it is the condition $R(\alpha)\equiv 0$ that actually determines the function $K(\alpha)$ (through equations (\ref{sum_S}),\,(\ref{sum_K}) above).  Provided the same $N$ is chosen in (\ref{sum_S}) and   (\ref{sum_xi}), this condition is actually identically satisfied for any $N$, as may be verified with some effort.
\vskip 2mm

\section{Appendix C: Numerical modelling}\label{App_Num}
\noindent Finite elements were used to determine the Stokes flow in the triply-connected fluid domain $\mathfrak{D}$ between the inner cylinders $\mathcal{C}_{1,2}$ and the outer  cylinder $\mathcal{C}_{0}$.  
On $\mathcal{C}_{1,2}$, the boundary conditions were  `no-slip', i.e.
$\left.\bfu\right|_{\{1,2\}}= \omega_{\{1,2\}} \, \bfe_z \wedge \bfn$  (with  unit  outward normal $\bfn$), and on $\mathcal{C}_{0}$,  the no-slip  condition 
$\left.\bfu\right|_{0}= 0$ was also imposed,  These are `essential'  boundary conditions in finite-element terms, in that
they are imposed explicitly on the solution $\bfu$.

The finite-element formulation involves writing the equations in
variational form; it is then required to  find $\bfu$ in $\calW_{\rm bc}$ and $\hat{p}$ in $\calQ$
such that
\begin{subequations}
 \label{Stokes_weak}
\begin{equation}
   2 \, \int_\mathfrak{D} {\bf e}(\bfu)\!:\!{\bf e}(\bfw) \,  \rmd \bfx
   - \int_\mathfrak{D}  \hat{p} \, \nabla\! \cdot \!\bfw \, \rmd \bfx = 0 \, , \qquad
   \forall \, \bfw \in \calW_0 \, ,
\label{Stokes_weak1}
\end{equation}
\vskip -5mm
\begin{equation}
 \int_\mathfrak{D}  q \, \nabla\! \cdot\! \bfu \, \rmd \bfx = 0 \, , \qquad \forall
 \, q \in \calQ \, ,
\end{equation}
\end{subequations} 
where $ {\bf e}(\bfu) $ is the rate-of-strain tensor, with cartesian
components  $e_{ij}$, and  $\calW_{\rm bc}\, ,$ $\calW_{0}$ and $\calQ$ are  suitably defined function spaces,
\begin{subequations} 
\begin{equation} 
\calW_{\rm bc} = \left\{ \bfw \in \left[H^1(\mathfrak{D} )\right]^2\,:\,
\left.\bfw\right|_{\{0,1,2\}}=\bfu|_{\{0,1,2\}}\right\} \, , 
\end{equation}
\vskip -5mm
\begin{equation} 
\calW_0 = \left\{ \bfw \in \left[H^1(\mathfrak{D})\right]^2\,:\,
\left.\bfw\right|_{\{0,1,2\}}=0\right\}  \quad {\rm and}\quad
\calQ = \left\{ q \in L^2(\mathfrak{D})\,:\, \langle q \rangle_\mathfrak{D} = 0 \right\} \, .
\end{equation}
\end{subequations} 
$\calQ$ consists of functions with zero mean on $\mathfrak{D} $; this allows for
the fact that the pressure is determined only up to an arbitrary
constant. These function spaces are then approximated with discrete spaces,
here by a triangular mesh, and with quadratic elements for the velocity and
affine elements for pressure. The resulting system was solved using {\sl
  FreeFem++} \citep{MR3043640}.

 Computing the streamfunction is not straightforward since the domain is not simply connected;  $\psi$ can be set to zero on only  one boundary:
 we set $\left.\psi\right|_{0}=0$ on $\mathcal{C}_0$.
The value of $\psi$ on $\mathcal{C}_1$ and $\mathcal{C}_2$ is not known \emph {a priori}. These missing boundary conditions can however be recovered from the solution of \eqref{Stokes_weak} as
\begin{equation} 
\psi_{\{1,2\}}= \pm \tfrac{1}{2}Q  \quad \mbox{with \ } \,
 Q = \int_{-\varepsilon}^{\varepsilon} v(x,0) \, \rmd x\, ,
\end{equation}
\vskip -5mm
\noindent and by solving
\vskip -5mm
\begin{equation} 
 \nabla ^2 \psi = -\left(\nabla \wedge \bfu\right) \cdot \bfe_z \, 
 \quad \mbox{with} \quad \left.\psi\right|_{\{0,1,2\}}=\psi_{\{0,1,2\}} \, ;
 \label{Stokes_stream2}
\end{equation}
or equivalently (but with higher-order shape functions) by solving $ \nabla ^4 \psi = 0$
with the additional boundary conditions $\left. \bfn \cdot \nabla \psi \right|_{\{1,2\}} = \mp 1$ respectively.

Some simulations using a stress-free outer boundary were also carried out for the co-rotating case. With $ \bfn$ and $ {\bf t}$ unit normal and tangent vectors on $\mathcal{C}_{0}$, the boundary conditions become  \begin{equation}
\left.\bfn \cdot \bfu \right|_{0} = 0\,, \quad \left. \bfn \cdot \bfe (\bfu) \cdot {\bf t} \right|_{0} = 0\,, \qquad\tn{on}\,\,\,\, \mathcal{C}_{0}\,,
\label{StressFree0}
\end{equation} 
and it is in general a non-trivial matter to implement these conditions.
In writing the variational form for this problem, the conditions on
$\left.\bfw\right|_{\!\{0\}}$ must be relaxed for both  $\calW_{\rm bc}$ and $\calW_0 \, .$
Since the full rate-of-strain tensor ${\bf e}$ was retained in the numerical formulation, the integration by parts that yields 
\eqref{Stokes_weak1} 
remains valid as stated with the boundary conditions
\eqref{StressFree0}.
The impenetrability condition is `essential', and  needs to be imposed strongly by a penalisation of the minimisation problem \eqref{Stokes_weak} with
\begin{equation}
\oint_{\mathcal{C}_{0}} ({\bf n}\cdot {\bf u})({\bf n}\cdot {\bf w}) \, \rmd s\, ,
\end{equation} 
where $s$ is arclength on $\mathcal{C}_{0}$, whereas the condition 
$\left. \bfn \cdot \bfe (\bfu) \cdot {\bf t} \right|_{0} = 0 $ is
`natural', following from the weak form \eqref{Stokes_weak}.

Finally, in order to resolve variables with sufficient accuracy over the full range of scales $[\varepsilon,R_0]$, and to reduce matrix sizes, an Uzawa splitting was employed in our largest simulations.

\vskip 5mm
\section*{Acknowledgments}
This research began during the Cambridge Lent Term, 2020, when ED was a Visiting Fellow at Trinity College, Cambridge; he wishes to express his gratitude to the College for the invitation that led to this visit and to the work recorded in this paper.


\clearpage

\

\vfill

\centerline{\Large \sc Supplementary material}

\vfill

\

\clearpage

\centerline{\large \sc Supplementary material}
\setcounter{equation}{0}
\setcounter{figure}{0}
\section{1. Sliced cylinders in contact, with counter-rotation ($\varepsilon<0$)} \label{Conveyor_counter}

\begin{figure}
\begin{center}
\includegraphics*[width=0.35\textwidth,  trim=0mm 0mm 0mm 0mm]{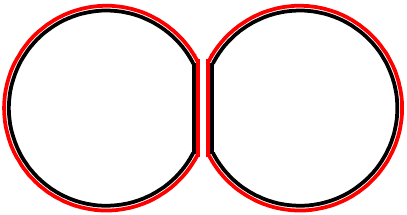} 
\end{center}
\caption{`Sliced counter-rotating cylinders' (black) with conveyor belts
  (red):   if the gap is reduced to zero, then the situation of
  `overlapping cylinders' with $\varepsilon<0$ is approximately
  realised.
}   \label{Sliced_cylinders}
\end{figure}

\noindent It is of some interest to  consider  the limit as $\varepsilon$ approaches zero from below.
The situation $\varepsilon<0$ corresponds to overlapping cylinders. With $\varepsilon_{1}=-\varepsilon >0$, the cylinders intersect at
$x=0,\, y=\pm (2\varepsilon_{1})^{1/2}$, and at a finite angle  $2\alpha \sim 2(2\varepsilon_{1})^{1/2}$. When $\varepsilon_{1}\ll 1$, and  $y^{2}> 2\varepsilon_1$, the fluid boundary in the lubrication region is at $x= \pm h(y)$ where now $h(y)=\thalf y^{2}-\varepsilon_1$. The situation could be approximated experimentally by using conveyor belts around two `sliced cylinders' and bringing the two flat faces together, as illustrated 
in figure \ref{Sliced_cylinders}(a). In biomechanics, it is a situation that could in principle be realised by ciliary propulsion of a  micro-organism. Equally, it is the situation first recognised by Hertz (1882), when two solid elastic bodies are in contact, one rolling on the other.

\begin{figure}
\begin{center}
\includegraphics*[width=\textwidth,  trim=0mm 0mm 0mm 0mm]{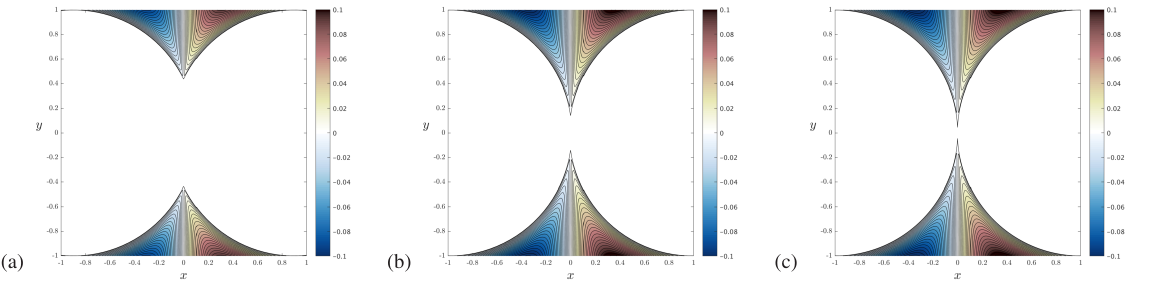} 
\end{center}
\caption{Streamlines for the flow (\ref{psi_overlap}) for  $\varepsilon_1 \equiv
  -\varepsilon = 10^{-1}\!, \,10^{-2}$ and
  $10^{-3}$.} 
  \label{2pannels_neg}
\end{figure}

For $0<\varepsilon_{1}\ll 1$, the velocity profile for $(2\varepsilon_{1})^{1/2}<|y|\ll 1 $ is still given by equation (19) of the main text, but the flux $Q$ is now zero so that $d\hat{p}/dy=3/h(y)^2 =12/(y^{2}-2\varepsilon_{1})^2$\,.
The pressure $\hat{p}(y)$ is now therefore given by
\be\label{pressure_sliced}
\hat{p}(y)=12
 \int{\frac{dy}{(y^{2}-2\varepsilon_{1})^2}}\,,
\ee
and the $y$-component of velocity is $v(x,y)= 3x^2/2h(y)^2 -\thalf$, with corresponding streamfunction 
\be \label{psi_overlap}
\psi _{L}(x,y) =x/2-x^3/2h(y)^2\,;
\ee
the streamlines $\psi _{L} =$ const.~are shown in figure \ref{2pannels_neg} for  $\varepsilon_1
= 10^{-1}, 10^{-2} \,\tn{and}\, 10^{-3}$. Obviously the description becomes more accurate as $\varepsilon_{1}\downarrow 0$ (i.e.~$\varepsilon\uparrow 0$).

\begin{figure}
  \centerline{\includegraphics*[width=0.35\textwidth,  trim=0mm 0mm 0mm 0mm]{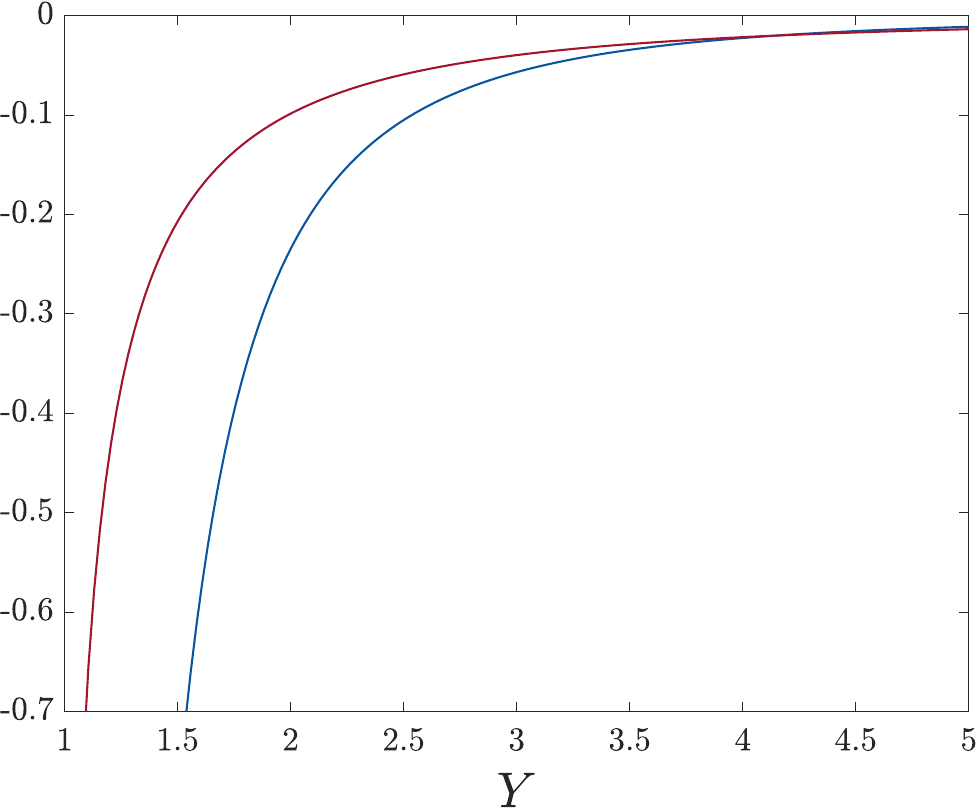}}
\caption{The pressure function  $-2Y(Y^2 -1)^{-1}+\log[(Y+1)/(Y-1)]$ (blue, equation (\ref{presover})) and the total vertical stress function  $1-\thalf Y \log[(Y+1)/(Y-1)]$ (red, equation (\ref{forcedens})). }
\label{Cyloverlap2}
\end{figure}

With $Y_{1}=y/(2\varepsilon_{1})^{1/2}>1$, the integral (\ref{pressure_sliced}) now gives
\be\label{presover}
\frac{ (2\varepsilon_{1})^{3/2}\hat{p}(y)}{3 } = 4\int{\frac{dY_{1}}{(Y_{1}^2-1)^2}}=\frac{-2Y_{1}}{Y_{1}^2 -1}+\log{\frac{Y_{1}+1}{Y_{1}-1}}\,.
\ee
This pressure function is shown by the blue curve in
 figure \ref{Cyloverlap2}.  The vertical component of the pressure force on the  boundary $x=h(y)$ for $y>  (2\varepsilon_{1})^{1/2}$, i.e. $Y_{1}>1$,  is then given by
\be\label{pressn}
- y \,\hat{p}(y)=\frac{3 }{\varepsilon_ {1}}\left( \frac{ Y_{1}^2}{Y_{1}^2 -1}-\frac{Y_{1}}{2}\log{\frac{Y_{1}+1}{Y_{1}-1}}\right)\,.
\ee


The viscous stress on the boundary $x=h(y)$ for $y>  (2\varepsilon_{1})^{1/2}$ is given by
\be\label{viscn}
\hat{\tau}(y)= -  \frac{\partial v}{\partial x}=-\frac{6 }{y^2 - 2\varepsilon_ {1}}=- \frac{3 }{\varepsilon_{ 1}(Y_{1}^2 -1)}.
\ee
It then follows from (\ref{pressn}) and  (\ref{viscn}) that the  vertical component of total stress on the boundary is 
\be\label{forcedens}
\hat{\sigma}(y) \equiv \hat{\tau}(y)-y\, \hat{p}(y) = 3    \varepsilon_{ 1}^{-1}\left(1-\frac{Y_{1}}{2} \log{\frac{Y_{1}\!+\!1}{Y_{1}\!-\!1}} \right)\,.
\ee
Note that the dominant terms of  order $(Y_{1}-1)^{-1}$ in (\ref{pressn}) and (\ref{viscn}) cancel, leaving the weaker logarithmic singularity at $Y_{1}=1$ in (\ref{forcedens}). (We shall see later [eqn.~(\ref{total_stress}) below]  that this singularity must result from the curvature of $\mathcal{C}_1$ and $\mathcal{C}_2$ away from the point of intersection $Y_{1}=1$). 

The total force from the curved parts of the composite body is now
\be\label{totalforcecomp}
\hat{F}_{yv} +\hat{F}_{yp}=4\int_{(2\varepsilon_{1})^{1/2}}^{\infty}(\hat{\tau}(y)-y\, \hat{p}(y) )\,dy =12  \sqrt{2}\,  \,\varepsilon_{1}^{-1/2}\int_{1}^{\infty}\!\left(1\!-\!\frac{Y_{1}}{2}\! \log{\frac{Y_{1}\!+\!1}{Y_{1}\!-\!1}} \right)dY_{1}\,.
\ee
The integrand is shown by the red curve in figure \ref{Cyloverlap2}; the singularity at $Y_{1}=1$ is integrable, and as $Y_{1}\rightarrow\infty$,
\be
1-\frac{Y_{1}}{2} \log{\frac{Y_{1}+1}{Y_{1}-1}} \sim -\tthird Y_{1}^{-2} \,.
\ee
The integral therefore converges; it actually evaluates to $-\thalf$, giving
\be\label{Force_curved}
\hat{F}_{yv} +\hat{F}_{yp}=-6\sqrt{2}\,  \,\varepsilon_{1}^{-1/2}\,.
\ee
\vskip 2mm \noindent
{\bf1.1 Distributed contact force}
\vskip 2mm \noindent
Again however, we have to consider whether the infinite difference in pressure between  $y=-(2\varepsilon_{1})^{1/2}$ and  $y=+(2\varepsilon_{1})^{1/2}$ may contribute to the total vertical force on the composite body.  To calculate this, consider the pressure  $p(y_1)$ at a small distance above $y=(2\varepsilon_{1})^{1/2}$ (i.e.~above  $Y_{1}=1$).  This is given by (\ref{presover}), and so (allowing also for the similar singularity at $Y_{1}=-1$) 
\bea \label{contact_force2}
\hat{F}_{c} &=&\lim_{Y_{1}\rightarrow 1}\int_{-h(y_1)}^{h(y_1)}[\hat{p}(-y_1)-\hat{p}(y_1)]\,\tn{d}x\nonumber \\ &=&\frac{6 }{(2\varepsilon_{1})^{1/2}}\lim_{Y_{1}\rightarrow 1}\left\{2Y_{1}-(Y_{1}^2-1)\log \left[\frac{Y_{1}+1}{Y_{1}-1}\right]\right\} =+6\sqrt{2}\,  \,\varepsilon_{1}^{-1/2}\,,
\eea
a value that again exactly compensates the contribution (\ref{Force_curved}); so again
\be \label{net_force_zero}
\hat{F}_{yv} +\hat{F}_{yp} +\hat{F}_{c} =0\,.
\ee
Here, the force $\hat{F}_{c}$ is due to the finite area of contact of the two sliced cylinders, and may therefore be appropriately described as a `distributed contact force'.

\begin{figure}
\begin{center}
  \includegraphics*[width=\textwidth,  trim=0mm 0mm 0mm 0mm]{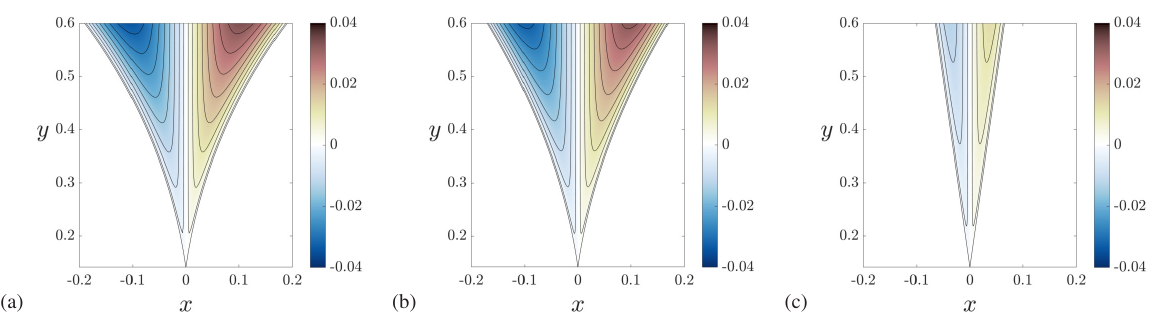} 
\end{center}
\caption{(a) Streamlines near the corner for $\varepsilon_1=10^{-2}$ ($\alpha = (2\varepsilon_{1})^{1/2}=0.1414$): (a) from the full numerics;
 (b) from lubrication theory (equation \ref {psi_overlap}), and (c) from the local similarity solution  (\ref{similarity}) and (\ref{ftheta}) (with boundaries at $\theta=\pm \alpha$ and the $y$-axis coinciding with $\theta=0$). The angle at the singular point $(0,\alpha)$ is $2\alpha$, the same in each case.}
  \label{Corner_flow}
\end{figure}

\vskip 2mm \noindent
{\bf1.2 Corner flow for overlapping cylinders}
\vskip 2mm \noindent
Since, when  $\varepsilon<0$, the cylinders intersect at a finite angle, we can alternatively consider the local similarity solution that exists very near the line of intersection, where the two boundaries can be treated as approximately plane.  Using polar coordinates $(r,\theta)$ such that the boundaries are (locally) at $\theta=\pm \alpha$, and with radial velocity $V=1$ on each boundary,  we may seek a similarity solution for the corner streamfunction $\psi_{C}(r,\theta)$ of the form
\be\label{similarity}
\psi_{C}(r,\theta)=r V f(\theta)\,,
\ee
with velocity components
\be
u_{r} = r^{-1}\partial \psi_{C}/\partial \theta =V  f'(\theta),\quad u_{ \theta} =- \partial \psi_{C}/\partial r=-V f(\theta)\,.
\ee
Since $u_{r}$ is evidently an even function of $\theta$, $f(\theta)$ must be odd; 
the boundary conditions are then satisfied if   
\be\label{bcs}
f(-\theta)=-f(\theta),\quad f(\alpha) = 0,\quad f'(\alpha) = -1.
\ee
The biharmonic equation $\nabla ^4 \psi=0$ has a solution of the form (\ref{similarity}) provided
\be
f(\theta)= A \sin\theta + B\cos\theta +C \theta \sin\theta +D \theta \cos\theta\,,
\ee
and the conditions (\ref{bcs}) determine the constants $A,B,C,D$, giving
\be\label{ftheta}
f(\theta)=\frac{\alpha\cos\alpha\sin\theta-\theta\cos\theta\sin\alpha}{\alpha-\sin\alpha\cos\alpha}\,.
\ee
 A  solution of similar form was first found for the `paint-scraper' problem by Taylor (1962) \cite{T62}. 

Figure \ref{Corner_flow} shows three panels, (a) showing the local streamline pattern from the full numerical solution for $\varepsilon_1=10^{-2}$, (b) showing the corresponding lubrication-theory solution (\ref {psi_overlap}), and (c) showing the corresponding  local similarity solution  (\ref{similarity}) and (\ref{ftheta}) for $\alpha = (2\varepsilon_{1})^{1/2}=0.1414$ (so that the angle at the singular point $(0,\alpha)$ is  $2\alpha$, the same in each case).

The pressure field $\hat{p}(r,\theta)$ is given by
\be\label{pressure_field_eqns}
\frac{\partial \hat{p}}{\partial r} =\left( \nabla^2 -\frac{1}{r^2}\right)u_{r} -\frac{2}{r^2}\,\frac{\partial u_{\theta}}{\partial \theta}\,,\quad \frac{1}{ \,r}\frac{\partial \hat{p}}{\partial\theta} =\left( \nabla^2 -\frac{1}{r^2}\right)u_{\theta} +\frac{2}{r^2}\,\frac{\partial u_{r}}{\partial \theta}\,.
\ee
Either of these equations leads to the result
\be
\hat{p}(r,\theta)  = -\frac{2   \cos\theta\sin\alpha}{r(\alpha-\cos\alpha\sin\alpha)}\,.
\ee
The tangential (radial) viscous stress on the boundary $\theta=\alpha$ is
\be
\hat{\tau}(r,\alpha)=-\frac{1}{r}\left.\frac{\partial u_{r}}{\partial \theta}\right |_{\theta=\alpha} =-\frac{2  \sin^{2} \alpha}{r(\alpha-\sin\alpha\cos\alpha)} \,.
\ee
The total stress on this boundary in the direction $\theta=0$ is now
\be\label{total_stress}
\hat{\sigma}(r,\alpha)\equiv \hat{\tau}(r,\alpha) \cos\alpha -\hat{p}(r,\alpha)\sin\alpha \equiv 0.
\ee
Thus the pressure force in the direction $\theta=0$  is exactly balanced by the viscous stress contribution [see the comment in parenthesis following equation (\ref{forcedens})].
\vskip 2mm \noindent
{\bf1.3 Corner contact force}
\vskip 2mm \noindent
But consider now, as before, the contact force $\hat{F}_{c}^{+}$ concentrated at the  corner.  This is obtained as
\be\label{Corner_contact}
\hat{F}_{c}^{+}=\lim_{r\rightarrow 0}\int_{-\alpha}^{\alpha}-\hat{p}(r,\theta)\,r\,\tn{d}\theta = \frac{4\sin^{2}\alpha}{\alpha-\sin\alpha\cos\alpha}\,.
\ee
\begin{figure}
(a)  {\includegraphics*[width=0.4\textwidth,  trim=0mm 0mm 0mm 0mm]{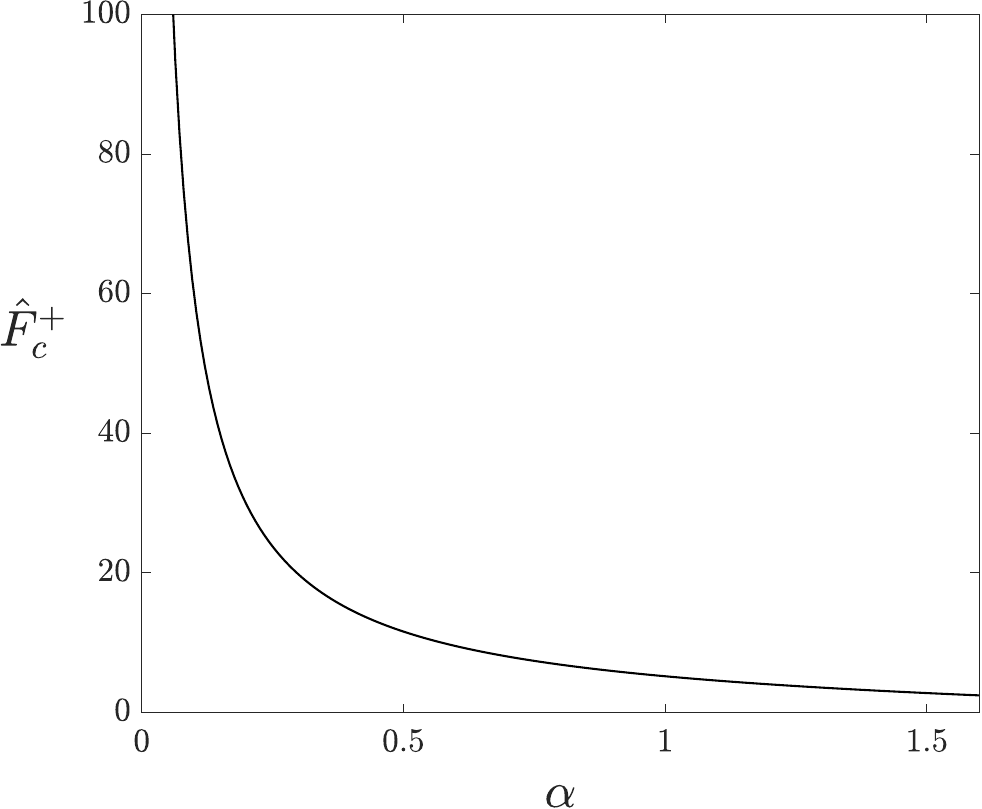}} \quad
  (b) \raisebox{1mm}{\includegraphics*[width=0.31\textwidth,  trim=0mm 0mm 0mm 0mm]{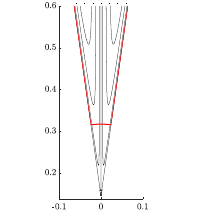}} 
\caption{(a) Contact force function $\hat{F}_{c}^{+}(\alpha)= 4\sin^{2}\alpha(\alpha-\sin\alpha\cos\alpha)^{-1}$  for  corner flow;   (b) contour, shaded red, for calculation of contact force; the net force on the straight segments $\theta=\pm \alpha$ in the direction $\theta=0$ is zero, whereas the suction force on the circular arc $r^2=x^2 +(y-\alpha)^2=$ cst.~is given by (\ref{Corner_contact}) in the limit $r\rightarrow 0$. }
  \label{Contact_force_corner}
\end{figure}
\!\!The function $\hat{F}_{c}^{+} (\alpha)$ is shown in figure \ref{Contact_force_corner}.
For small $\alpha$, this  suction force has the asymptotic form $\hat{F}_{c}^{+} \sim 6/\alpha -4\alpha/5 +\tn{O}(\alpha^3)$, or, setting $\alpha=(2 \varepsilon_{1})^{1/2}$, 
\be\label{corner_contact_force}
\hat{F}_{c}^{+} \sim 3\sqrt{2}\,\varepsilon_{1}^{-1/2}\,,
\ee
consistent with the result (\ref{contact_force2}) (given that here only the upper singularity is considered; an equal contribution comes from the lower singularity).  The fact that this contact force is not compensated by contributions from the straight boundaries $\theta =\pm \alpha$ is a clear indication that it is the curvature away from the singularity that leads to the non-zero compensating contribution evident in (\ref{net_force_zero}).  

\section{2. The co-rotating `conveyor-belt' situation when $\varepsilon<0$ }\label{Conveyor_corot}
\begin{figure}
\begin{center}
\includegraphics*[width=0.3\textwidth,  trim=0mm 0mm 0mm 0mm]{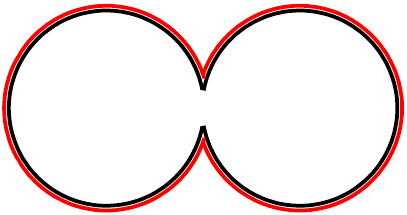} 
\end{center}
\caption{Co-rotating cylinders for $\varepsilon <0$ (black) simulated with a single conveyor belt (red).} 
  \label{Sliced_cylinders_corot}
\end{figure}
\begin{figure}
\centerline{\includegraphics*[width=\textwidth,  trim=0mm 0mm 0mm 0mm]{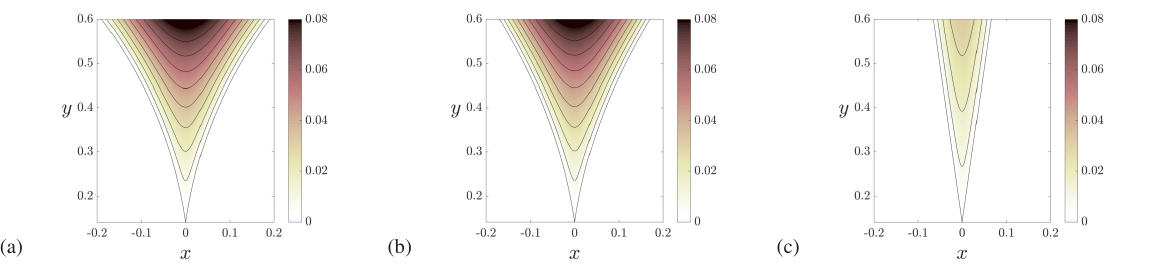}}  
\caption{Flow near the singular point for the co-rotating
  situation when  $\varepsilon_{1} (=-\varepsilon)=0.01$: (a) 
  the full numerics with no-slip boundary conditions; (b) the lubrication
  flow streamlines  $\psi_{L}(x,y)=$ const; (c) the corner flow streamlines
  $\psi_{C}(r,\theta)=$ const.~with $\psi_{C}(r,\theta)$ given by
  (\ref{corner_str_corot}) with the same angle $2\alpha=0.2826\approx 16^{\tn{o}}$ at the tip
 and with much-reduced vertical scale. } 
\label{Corner_corot}
\end{figure}
\vskip 2mm \noindent
As for the counter-rotating situation, we may consider the case $\varepsilon<0$, which, for co-rotating cylinders, may in principle be generated by a single conveyor belt, as illustrated in figure \ref{Sliced_cylinders_corot}.  Setting $\varepsilon_{1}=-\varepsilon >0$, the lubrication solution of \S 2 is applicable when $0<\varepsilon_{1}\ll 1$ in the fluid region 
$(2\varepsilon_{1})^{1/2}<|y|\ll 1$.  The streamlines in this region for $\varepsilon_{1}=0.01$ for the full numerical solution and for the corresponding lubrication solution  are shown in figure \ref{Corner_corot}(a,b);  the agreement is very close. For this value of  $\varepsilon_{1}$, the boundaries intersect at the point $(0, y_{0})$ at angle $2\alpha$, where $y_{0}=\alpha=\cosh^{-1}[1.01]\approx 0.1413 \,(\approx 8^{\tn{o}})$.

The corner flow  is now given by the streamfunction 
\be\label{corner_str_corot}
\psi_{C}(r,\theta)=\frac{r(\alpha\cos\theta\sin\alpha -\theta\sin\theta\cos\alpha)}{\alpha +\cos\alpha\sin\alpha}\,,
\ee
with associated velocity components
\be
u_{r}(\theta)\!=\!-\frac{\sin\theta(\alpha\sin\alpha \!+\!\cos\alpha)\!+\!\theta\cos\theta\cos\alpha }{\alpha +\sin\alpha\cos\alpha}\,,\,\,
u_{\theta}(\theta)\!=\!\frac{\theta\cos\alpha\sin\theta\!-\!\alpha\cos\theta\sin\alpha }{\alpha +\sin\alpha\cos\alpha}\,,
\ee
satisfying $u_{\theta}(\theta)=0, u_{r}(\theta)=\pm 1$ on $\theta=\mp\alpha$ respectively. The streamlines $\psi_{C}(r,\theta)=$ const.~are shown in  figure \ref{Corner_corot}(c), for angle $\alpha= 0.1413$. The pressure field for this flow  is given by
\be
\hat{p}(r,\theta)= -\frac{2\sin\theta\cos\alpha}{r(\alpha+\sin\alpha\cos\alpha)}\,.
\ee
The total vertical force acting on the boundary planes is zero by symmetry.

\end{document}